\PassOptionsToPackage{dvipsnames,svgnames,table}{xcolor}

\documentclass[
    aps,
    prd,
    a4paper,
    twocolumn,
    amsmath,
    showpacs,
    superscriptaddress,
    nofootinbib,
    preprintnumbers
]{revtex4-1}

\pdfoutput=1

\usepackage{amssymb,amsmath,latexsym,mathrsfs,bm}
\usepackage[sort&compress]{natbib}

\usepackage{graphicx}
\usepackage{subfigure}

\usepackage{epsfig}

\usepackage{varioref,xr-hyper}

\usepackage{multirow}
\usepackage{array}

\usepackage{orcidlink}
\usepackage{wasysym}
\usepackage{comment}

\usepackage[utf8]{inputenc}
\usepackage[T1]{fontenc}

\usepackage[normalem]{ulem}

\usepackage{hyperref}

\hypersetup{
    colorlinks=true,
    linkcolor=blue,
    citecolor=red,
    urlcolor=magenta
}

\newcommand{\bea}{\begin{eqnarray}}
\newcommand{\eea}{\end{eqnarray}}

\newcommand{\mnu}{\sum m_{\nu,{\rm eff}}}

\newcommand{\be}{\begin{equation}}
\newcommand{\ee}{\end{equation}}

\begin{document}


\title{When One-Parameter Dark Energy Makes Neutrinos Physical Again}

\author{Weiqiang Yang\orcidlink{0000-0002-6486-6765}}
\email{d11102004@163.com}
\affiliation{Department of Physics, Liaoning Normal University, Dalian, 116029, P. R.
China}

\author{Eleonora Di Valentino\orcidlink{0000-0001-8408-6961}}
\email{e.divalentino@sheffield.ac.uk}
\affiliation{School of Mathematical and Physical Sciences, University of Sheffield, Hounsfield Road, Sheffield S3 7RH, United Kingdom}

\author{Eric V.\ Linder\orcidlink{0000-0001-5536-9241}}
\email{evlinder@lbl.gov}
\affiliation{Berkeley Center for Cosmological Physics \& Berkeley Lab,
University of California,\\ Berkeley, CA 94720, USA}

\author{Sibo Zhang\orcidlink{0009-0003-1425-540X}}
\email{sbzhang02@163.com}
\affiliation{Department of Physics, Liaoning Normal University, Dalian, 116029, People's Republic of China}

\author{Supriya Pan\orcidlink{0000-0001-9500-401X}}
\email{supriya.maths@presiuniv.ac.in}
\affiliation{Department of Mathematics, Presidency University, 86/1 College Street, Kolkata 700073, India} 
\affiliation{Institute of Systems Science, 
Durban University of Technology, Durban 4000, Republic of South Africa}


\begin{abstract}
A puzzling implication of current data interpreted in the $\Lambda$-Cold Dark Matter cosmology is the preference for a negative sum of neutrino masses. Moving to $w_0w_a$CDM brings an appreciable fraction of the neutrino mass posterior back to positive values, while the constant equation-of-state dark energy case $w$CDM does not. We investigate a variety of  one-parameter  dynamical dark energy equations of state (DE EoS) whose redshift evolution is controlled by a single free parameter,
each representing distinct physical properties,
to understand whether a two-parameter DE EoS is required to bring the neutrino mass into the positive region. The conclusion is that certain one-parameter DE EoS can suffice, implying that the data are pointing toward physical characteristics rather than a broad degeneracy.  This behavior effectively lowers the dark energy density at high redshift, allowing the sum of neutrino masses to shift toward the physical region. The required characteristics are identified as phantom dark energy at high redshift, crossing $w=-1$ at lower redshift. 
\end{abstract}

\maketitle
\section{Introduction}

The Dark Energy Spectroscopic Instrument (DESI), combining its data with other cosmic distance measurements, has recently reported intriguing indications for dynamical dark energy (DDE) from its first and second data releases~\cite{DESI:2024mwx, DESI:2025zgx}. These results challenge the standard $\Lambda$-Cold Dark Matter ($\Lambda$CDM) cosmological model, in which dark energy (DE) is described by a cosmological constant corresponding to vacuum energy with a constant equation-of-state (EoS) $w=-1$. 
At the same time, when working {\it within\/} the $\Lambda$CDM model, the results indicate the posterior for the sum of neutrino masses to lie substantially at  unphysically negative values. We aim to investigate to what extent these two indications are connected. 

According to the latest DESI results~\cite{DESI:2025zgx}, the combination of cosmic microwave background (CMB) data with baryon acoustic oscillations (BAO) measurements from DESI DR2 provides evidence not only for DDE over $\Lambda$CDM, but that the dark energy dynamics evolved from $w<-1$ to $w>-1$. 
This has been indicated not only by the widely used $w_0w_a$ parametrization of the DE
EoS~\cite{Chevallier:2000qy, Linder:2002et}, where 
$w(a)=w_0+w_a(1-a)$, but by form independent methods such 
as orthogonal polynomials, independent redshift binning, 
and Gaussian process regression~\cite{DESI:2025fii}. 
Depending on the Type Ia supernovae (SNIa) compilation adopted, the significance initially ranged from $2.8\sigma$ to $4.2\sigma$~\cite{DESI:2025zgx}. More recently, improved calibrations of supernovae datasets, including DES-Dovekie~\cite{DES:2025sig} and Union3.1~\cite{Hoyt:2026fve}, 
have brought the different SNIa compilations into better agreement. Using these recalibrated datasets, the evidence for DDE is found to lie in the narrower range $3.2\sigma$--$3.4\sigma$~\cite{DES:2025sig,Hoyt:2026fve}, strengthening the robustness of the result and reducing the dependence on the specific supernovae sample used. These findings have renewed interest in exploring extensions of $\Lambda$CDM and in assessing the physical origin of the possible DE 
dynamics~\cite{Cortes:2024lgw,Shlivko:2024llw,Luongo:2024fww,Gialamas:2024lyw,Dinda:2024kjf,Najafi:2024qzm,Wang:2024dka,Ye:2024ywg,Tada:2024znt,
Carloni:2024zpl,Chan-GyungPark:2024mlx,Bhattacharya:2024hep,Ramadan:2024kmn,Pourojaghi:2024tmw,Giare:2024gpk,Reboucas:2024smm,Giare:2024ocw,Chan-GyungPark:2024brx,Li:2024qus,Jiang:2024viw,Jiang:2024xnu,RoyChoudhury:2024wri,Li:2025cxn,Paliathanasis:2025dcr,Wolf:2025jlc,Shajib:2025tpd,Giare:2025pzu,Paliathanasis:2025hjw,Chaussidon:2025npr,Kessler:2025kju,Pang:2025lvh,Roy:2024kni,RoyChoudhury:2025dhe,Paliathanasis:2025cuc,Scherer:2025esj,Liu:2025mub,Teixeira:2025czm,Santos:2025wiv,Specogna:2025guo,Sabogal:2025jbo,Giare:2025ath,Cheng:2025lod,Herold:2025hkb,Cheng:2025hug,Ozulker:2025ehg,Lee:2025pzo,Ormondroyd:2025iaf,Li:2025dwz,Silva:2025twg,Ishak:2025cay,Fazzari:2025lzd,Smith:2025icl,Zhang:2025lam,Cheng:2025yue,Zhou:2025nkb,Li:2025vuh,Cheng:2025yue,Ladeira:2026jne,Paliathanasis:2026ymi,Li:2026xaz,Gokcen:2026pkq,Feng:2026pzs,Luciano:2026vhm}.  In this article, we focus on what we refer to as one-parameter dynamical dark energy models, namely the parametrizations whose redshift evolution is governed by a single free parameter. 

At the same time, DESI has provided some of the strongest cosmological constraints on the sum of the neutrino masses, $\sum m_\nu$. Assuming the physical prior $\sum m_\nu \geq 0$, the DESI DR2 analysis reports the following $95\%$ upper limits from the dataset combination DESI DR2 BAO + CMB + PantheonPlus~\cite{Elbers:2025vlz}: 
\begin{align}
\sum m_\nu &< 0.0704\,\mathrm{eV} \quad (\Lambda\mathrm{CDM}+\sum m_\nu), \\
\sum m_\nu &< 0.0653\,\mathrm{eV} \quad (w\mathrm{CDM}+\sum m_\nu), \\
\sum m_\nu &< 0.117\,\mathrm{eV} \quad (w_0w_a\mathrm{CDM}+\sum m_\nu).
\end{align}
While these limits are consistent with the expectation that neutrino masses must be 
bounded from below by neutrino oscillation experiments, several recent cosmological analyses have pointed out an intriguing feature: when the prior $\sum m_\nu \geq 0$ is relaxed, the posterior distribution may peak at negative values of the neutrino mass parameter~\cite{Craig:2024tky,Green:2024xbb,Elbers:2025vlz,Pulido-Hernandez:2026hcs}. 

Although a negative neutrino mass is unphysical, a posterior distribution that peaks at negative values
may arise as an effective phenomenological signal indicating a mismatch between the assumed cosmological model and the data. In the past, some work has alternately linked a preference for $\sum m_\nu <0$ to parameter degeneracies, residual tensions between different cosmological datasets, or enhanced lensing or clustering amplitudes~\cite{Elbers:2024sha,Naredo-Tuero:2024sgf}. Understanding the origin of this behaviour remains an open problem in cosmology.

Interestingly, recent studies have shown that allowing for DDE can alleviate the preference for negative neutrino masses. For instance, analyses adopting the $w_0w_a$ parametrization have found that the introduction of DE  
dynamics can shift the posterior of $\sum m_\nu$ toward positive values and relax the tension with neutrino oscillation bounds~\cite{Elbers:2024sha}. This raises an important question: is the apparent resolution of the negative neutrino mass preference simply a consequence of increased parameter freedom in models with multiple DE parameters, or does it point toward specific physical properties of the DE sector?

Addressing this question requires exploring minimal extensions of $\Lambda$CDM in which DE 
dynamics are controlled by a reduced number of parameters. From this perspective, DDE parametrizations with a single additional parameter are particularly appealing. Besides providing a physically motivated description of DE 
evolution, such models avoid the excessive parameter freedom of multi-parameter frameworks and are less susceptible to penalties from Bayesian model comparison.
In this work we investigate the behaviour of the total neutrino mass within a set of well-motivated one-parameter DDE models. Specifically, we consider thawing models~\cite{dePutter:2008wt,Linder:2015zxa}, generalized thawers~\cite{Linder:2007wa}, the mirage dark energy 
model~\cite{Linder:2007ka}, and the generalized emergent dark energy 
model~\cite{Li:2020ybr,Yang:2021eud}. These parametrizations represent distinct physical behaviours of the DE 
sector while maintaining minimal parameter freedom. By analysing the constraints on $\sum m_\nu$ within these frameworks using a combination of recent cosmological datasets, we aim to determine whether the preference for negative neutrino masses is purely a degeneracy effect associated with specific parametrizations, or whether it carries information about the underlying physics of DE. 

The article is structured as follows. In Section~\ref{sec-2} we briefly describe the cosmological equations for a general 
DE 
model and introduce the one-parameter DDE parametrizations considered in this work. In Section~\ref{sec-data} we present the observational datasets and the numerical methodology adopted to constrain the models. In Section~\ref{sec-results} we discuss the results obtained from the data analysis. Finally, in Section~\ref{sec-summary} we summarize the main conclusions of this study.

\section{One parameter Dynamical Dark Energy}
\label{sec-2}

Starting with the spatially flat Friedmann-Lema\^itre-Robertson-Walker (FLRW) line element and General Relativity describing the universe's gravitational sector, the cosmic expansion rate is given by the Hubble parameter,
\begin{eqnarray}\label{Hubble-expansion}
    \frac{H^2}{H_0^2} = \Omega_{m0} a^{-3} + \Omega_{r0} a^{-4} + \Omega_{\nu,0} \frac{\rho_{\nu}}{\rho_{\nu,0}} \nonumber\\+ \Omega_{\rm de,0} \exp\left(3\int_{a}^{1} \frac{1+w (a')}{a'}~da' \right)
\end{eqnarray}
where we have assumed that the fluids, namely pressure-less matter (cold dark matter plus baryons), radiation, neutrinos, and DE under consideration do not interact (at the non-gravitational level) with each other. 
In Eq.~(\ref{Hubble-expansion}), $\Omega_{m0}$, $\Omega_{r0}$, $\Omega_{\rm de,0}$, and $\Omega_{\nu,0}$ are respectively the density parameters of matter, radiation, DE, and neutrinos at the present epoch, 
and $\Omega_{\nu,0}$ relates to the sum of the neutrino masses $\sum m_{\nu}$ as $\Omega_{\nu,0} {h}^2 = \sum m_{\nu}/(93.14~{\rm eV})$, 
in which ${h} = H_0$/(100\, km/s/Mpc).
The function $w(a)$ denotes the pressure-to-energy-density, or EoS,
ratio of DE.

We focus on one-parameter DE EoS models, since DESI~\cite{DESI:2025zgx} already showed that positive 
neutrino mass has reasonable support for the two-parameter $w_0w_a$ DE 
model. 
To investigate the reason why the $w_0w_a$ cosmology enables 
positive neutrino mass we consider: 1) Are two parameters 
for the DE EoS  
a key element?, 2) Is rapid 
evolution (e.g.\ large amplitude $w_a$) at recent times 
($z\lesssim0.5$) key?, 3) Is lower 
DE 
density 
(i.e.\ $w<-1$) at higher redshift ($z\gtrsim1$) key?

To test these possibilities we use one-parameter 
 DE 
models with low-redshift rapid evolution and high-redshift 
low energy density (mirage dark energy case), 
DE 
with recent rapid evolution but not low density at high 
redshift (generalized thawing cases), 
DE 
with low 
density at high redshift but not rapid evolution recently 
(GEDE case), and DE 
with a smaller amount of rapid 
evolution recently and a smaller lowering of the energy density 
at high redshift (conventional, sometimes called calibrated, 
thawing case; also as a cross-check on one of the generalized 
thawing cases). We emphasize that these are not randomly chosen 
parametrizations, but each is used to focus on a specific 
physical characteristic.
 
{\bf Thawing dark energy} 
acts like a frozen 
scalar field, hence a cosmological constant, in 
the matter-dominated era and gradually moves 
away from $w=-1$ at later times (e.g.\ when the steepness 
of the potential overcomes the diminishing Hubble friction). 
A simple one-parameter EoS that reproduces the observables 
(distance and Hubble parameter) to $\sim0.1\%$ for a wide class of 
thawers, including potentials linear, quadratic, and quartic 
in the field, and hilltop/PNGB (pseudo-Nambu Goldstone boson) 
cases~\cite{dePutter:2008wt,Linder:2015zxa}, is 
\be 
w_a=-1.58(1+w_0)\ . 
\ee

\begin{figure}
    \centering
    \includegraphics[width=0.45\textwidth]{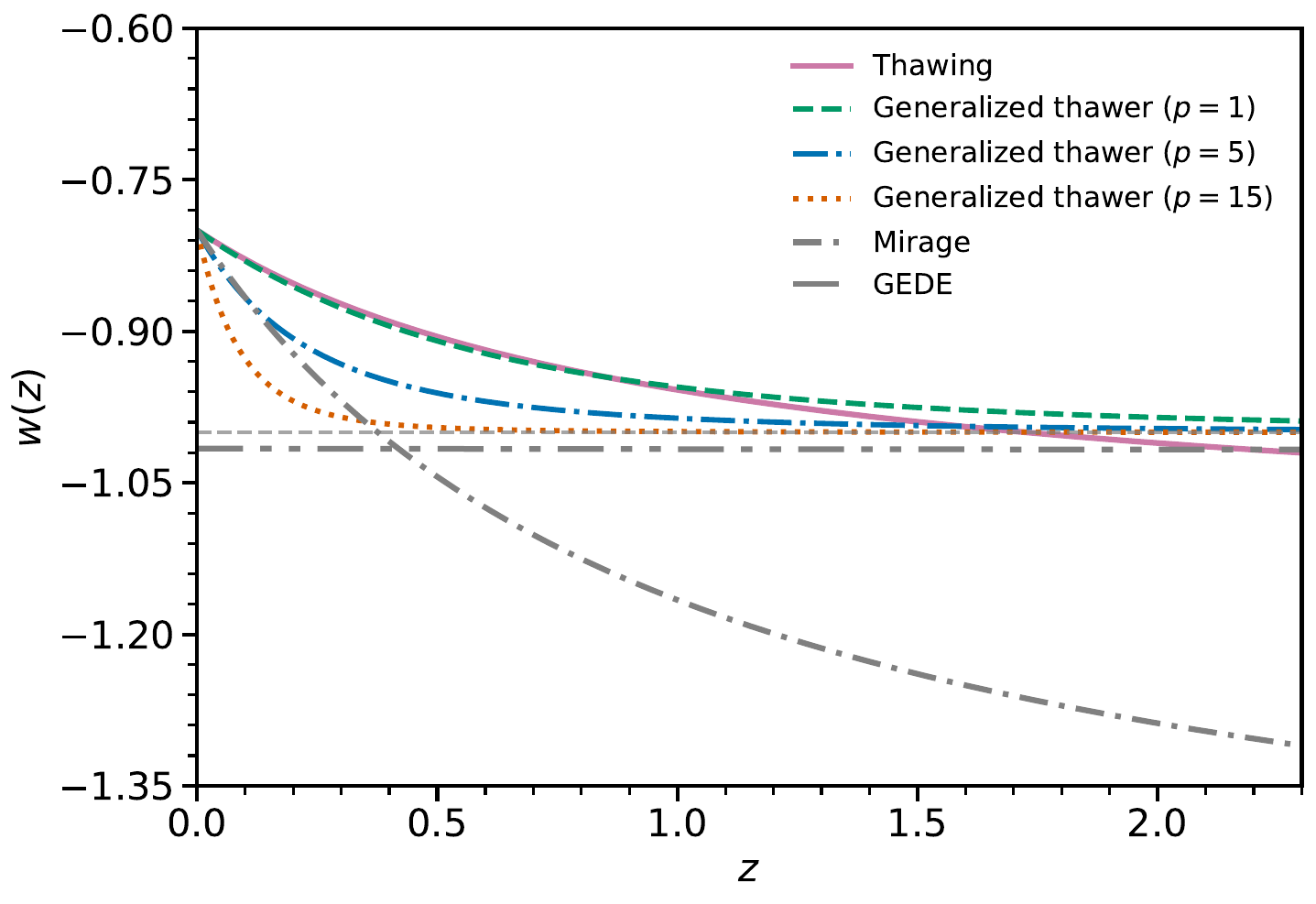}
    \caption{Evolution of $w(z)$ for the {\bf thawing}, generalized thawing models for three values of $p$, {\bf mirage}, and {\bf generalized emergent} dark energy models. In this figure, we fix $w_0=-0.8$ for all dark energy models except the generalized emergent dark energy model, for which we adopt $\delta =0.05$. In the generalized thawing model, the effect of the parameter $p$ on the steepness of the evolution is clearly visible. The horizontal dotted line corresponds to $w=-1$.}
    \label{fig:eos}
\end{figure}

{\bf Generalized thawers} can have a more rapid late-time 
evolution than standard thawers. This has gained interest 
with the DESI results, and as a way of exploring whether a 
rapid evolution near the present could obviate the need for 
the EoS 
to cross $w=-1$ in the past. While results show 
that it cannot (see, for example, the model-independent reconstructions 
and Table 3 in~\cite{DESI:2025fii}), 
such models are of interest 
in testing how the rapid evolution affects the neutrino mass 
constraint. The one-parameter EoS 
for generalized thawers is~\cite{Linder:2007wa}
\be 
1+w(a) =(1+w_0)a^3 \left(\frac{3}{1+2a^3} \right)^{1-p/3}\ . 
\ee 
Note that $w\to-1$ in the past, and deviates from it later 
as $a^3$ initially, the criterion for the thawing class, 
but evolves away from $w=-1$ as $a^p$ at asymptotically late 
times. The EoS always has $w\ge-1$, and the larger $p$ is, the 
later and more rapid is the deviation from $w=-1$; see Figure~\ref{fig:eos}.  

Physically, a way to obtain this ``supercharged'' 
evolution is a hilltop/PNGB field that starts very close to 
the false maximum and has a very steep slope (see, for example,~\cite{Shlivko:2024llw}). In fact, this has two fine tunings: the 
symmetry-breaking energy scale $f$ for the PNGB model requires 
$f\ll M_{Pl}$ (giving the steepness and hence late rapid 
evolution), and further the initial condition on the field 
must be exponentially close to the peak, $\phi_i/f\sim e^{-M_{\rm Pl}/f}$~\cite{dePutter:2008wt}, to give $w\approx-1$ even as 
the DE 
density fraction becomes appreciable, until the 
rapid evolution begins. Nevertheless, we use it to explore the 
consequences for neutrino mass. We adopt either $p=1$, which 
gives results extremely similar to the standard thawing case 
above, and $p=5$ and $p=15$ to quantify the effects of more rapid 
evolution.

{\bf Mirage dark energy}
~\cite{Linder:2007ka}, given by 
\be 
w_a=-3.66(1+w_0)\ , 
\ee 
preserves the 
distance to last scattering of the CMB, and DESI results show 
that it runs nearly along the main uncertainty direction of the $w_0w_a$ 
constraints from combined observational data. Thus we expect 
it to provide a good fit to the data here, and we can investigate 
the effect of this EoS 
having one less degree of freedom on the 
neutrino mass constraint. 

{\bf Generalized emergent dark energy (GEDE)}~\cite{Li:2020ybr,Yang:2021eud} 
is a phenomenological model where dark energy rises from a 
lower energy density than $\Lambda$CDM at high redshift ($w<-1$) 
and transitions toward $w=-1$ at later times. Its form is 
\be 
w(a)=-1-\frac{\delta}{3}\,\left(1-\tanh\left[\delta\,\ln\left(\frac{a}{a_e}\right)\right]\right)\ , 
\ee 
where the transition scale factor is defined through 
$\rho_{\rm de}(a_e)=\rho_m(a_e)$, leaving the transition width 
$\delta$ as the single parameter. (Note that in~\cite{Li:2020ybr,Yang:2021eud} they 
use $\Delta=\delta\,\ln 10$ and $1+z_t=a_e^{-1}$.) At high 
redshift, $a\ll a_e$, $w\to-1-2\delta/3$, while for $a\gg a_e$, 
$w\to -1$. For $\delta>0$ the high-redshift dark energy density 
is suppressed compared to $\Lambda$CDM, which 
is what we focus on. 
(Thus the prior of $\Delta\in[-3,10]$ used in 
some previous analyses here becomes $\delta\in[0,4.34]$.)

While $w(a)$ gives the background evolution for the dark energy, 
we also need  
to consider the 
density perturbation of the DE fluid and its velocity divergence. We restrict ourselves to the linear perturbation regime and consider the perturbed FLRW line element in the synchronous gauge given by $ds^2 = a^2(\tau) \Big[ 
- d\tau^2 + (\delta_{ij} + h_{ij})dx^{i} dx^{j} \, \Big],$ where the scale factor $a(\tau)$ is expressed in terms of the conformal time $\tau$, and $\delta_{ij}$ and $h_{ij}$ respectively refer to the unperturbed and perturbed spatial parts of the metric tensor.   
Now introducing $\delta_i = \delta \rho_i / \rho_i$ as the dimensionless density perturbation for the fluid component $i$ and 
$\theta_i$ 
as the divergence of the $i$-th fluid velocity, one can find the evolution equations of the above two quantities in Fourier space as~\cite{Ma:1995ey}:
\begin{eqnarray}
\delta'_{i} & = & - (1+ w_{i})\, \left(\theta_{i}+ \frac{h'}{2}\right) - 
3\mathcal{H}\left(\frac{\delta P_i}{\delta \rho_i} - w_{i} \right)\delta_i \nonumber \\
& & -  9 \mathcal{H}^2\left(\frac{\delta P_i}{\delta \rho_i} - c^2_{a,i} \right) (1+w_i) 
\frac{\theta_i}
{{\kappa}^2}, \label{per1} \\
\theta'_{i} & = & - \mathcal{H} \left(1- 3 \frac{\delta P_i}{\delta
\rho_i}\right)\theta_{i} 
+ \frac{\delta P_i/\delta \rho_i}{1+w_{i}}\, {\kappa}^2\, \delta_{i} 
-{\kappa}^2\sigma_i,\label{per2}
\end{eqnarray}
where the derivative of a variable with respect to the conformal time $\tau$ is denoted by a prime, $\mathcal{H}$ refers to the conformal Hubble parameter, $h$ (only here) corresponds to the synchronous gauge metric perturbation, $\kappa$ refers to the wavenumber in Fourier space, and $\sigma_i$ is the anisotropic stress of the $i$-th fluid, which in this article we set $\sigma_i = 0$. 
Moreover, in Eqs.~(\ref{per1}) and (\ref{per2}) $\delta P_i/\delta\rho_i$ is the rest-frame square of the sound speed of the $i$-th fluid and
$c^2_{a,i}=w_i-w_i'/[3\mathcal{H}(1+w_i)]$
is the adiabatic sound speed. During the statistical analysis $c^2_{\rm s,de}$ has been set to unity, as usually assumed for minimally coupled scalar field models, and for the matter sector we set $c^2_{\rm s,m}=0$.

\section{Observational data and methodology}
\label{sec-data}

This section summarizes the observational datasets and the numerical methodology used to constrain the parameter space of the proposed DE scenarios. For the one-parameter DE cosmologies we investigate two scenarios: one where the sum of the  neutrino masses $\sum m_{\nu,{\rm eff}}$ has been fixed to $0.06$ eV, and the other where $\mnu$ is kept as a free parameter. We follow Appendix A of \cite{Elbers:2025vlz} in treating $\mnu$ for negative values. 
This does not affect the high-redshift effective number of neutrino species $N_{\rm eff}$, which is fixed to $3.04$.

In the first set where $\sum m_{\nu,{\rm eff}}$ has been kept fixed, each cosmological model contains seven free parameters, 
\be 
{\bm \theta}=\Bigl\{\Omega_{b}h^2, \Omega_{c}h^2, 100\theta_{\rm MC}, \tau, n_{s}, \log[10^{10}A_{s}], w_0\Bigr\}\,, 
\ee 
while there are eight when $\sum m_{\nu,{\rm eff}}$ is free to vary, 
\be 
{\bm \theta_\nu}=\Bigl\{{\bm \theta},\,\mnu\Bigr\}\ . 
\ee
In the GEDE case $w_0$ is replaced by $\delta$. 
Apart from the one DE parameter ($w_0$ or $\delta$) and the total neutrino mass $\sum m_{\nu,{\rm eff}}$, the remaining parameters are: physical baryon density $\Omega_{\rm b} h^2$, physical cold dark matter density $\Omega_{\rm c} h^2$, angular scale of the sound horizon at recombination $\theta_{\rm MC}$, optical depth to reionization $\tau$, and the spectral index $n_{\mathrm{s}}$ and amplitude $A_{\mathrm{s}}$ of the primordial scalar perturbations.

The observational datasets used are: 

\begin{itemize}
	
	\item {\bf Cosmic Microwave Background Radiation}: Cosmic Microwave Background (CMB) data from Planck 2018~\cite{Planck:2018vyg,Planck:2019nip} have been used. Specifically, we consider the CMB temperature and polarization angular power spectra {\it plikTTTEEE+lowl+lowE}. We shall label this dataset as {\bf CMB}. 
	
	\item {\bf Baryon Acoustic 
    Oscillations} (BAO) scale measurements from the second data release (DR2) of the DESI survey~\cite{DESI:2025zgx}, providing cosmological distance constraints. In the following, we shall denote the BAO dataset as {\bf DESI}. 
	
	\item {\bf Type Ia supernovae (SNIa)} datasets from one of four sources: 
    \begin{enumerate}
    \item {\bf PantheonPlus} spectroscopic supernovae compilation~\cite{Scolnic:2021amr};

\item  {\bf Union3} spectroscopic supernovae compilation analyzed with the UNITY1.5 Bayesian hierarchical framework~\cite{Rubin:2023ovl}; 

\item {\bf Union 3.1} compilation analyzed with the updated UNITY1.8 framework~\cite{Hoyt:2026fve,Rubin:2026qdt}; 

\item  {\bf DES-Dovekie} reanalysis of the photometrically classified supernovae from the Dark Energy Survey~\cite{DES:2025sig}.
        
    \end{enumerate}

\end{itemize}

We constrained the parameter space given the data through Markov Chain Monte Carlo (MCMC) analyses using {\tt CLASS}~\cite{Blas:2011rf} and {\tt Cobaya}~\cite{Torrado:2020dgo}. The convergence of the MCMC chains is tested using the Gelman-Rubin parameter $R - 1$~\cite{Gelman:1992zz}, and we take the results from each chain when $R - 1 < 0.02$ is satisfied. 
In Table~\ref{tab:priors}, we list the uniform priors on the cosmological parameters.

\begin{table}
	\begin{center}
		\renewcommand{\arraystretch}{1.4}
        		\begin{tabular}{|c|c|}
			\hline
			\textbf{Parameter}           & \textbf{Prior}\\
			\hline\hline
			$\Omega_{\rm b} h^2$             & $[0.005,0.1]$ \\
			$\Omega_{\rm c} h^2$             & $[0.001,0.99]$\\
			$\tau$                       & $[0.01,0.8]$\\
			$n_{\rm s}$                        & $[0.8, 1.2]$\\
			\ \ $\log[10^{10}A_{\rm s}]$\ \          & \ \ $[1.61,3.91]$ \ \ \\
			$100\theta_{\rm MC}$             & $[0.5,10]$\\
			$\sum m_{\nu,{\rm eff}}$                      & $[-1.5,1.5]$\\			
			$w_0$                        & $[-3,1]$\\
            $\delta$   & $[0, 4.34]$ \\
			\hline
		\end{tabular}
	\end{center}
	\caption{Uniform priors used for the cosmological parameters in the statistical analysis.
    }
	\label{tab:priors}
\end{table}

We also examine the statistical viability of the proposed DE models through model comparison statistics. One is the usual $\chi^2_{\rm min}$ statistic and the other uses Bayesian evidence. We compute $\Delta \chi^2_{\rm min}$ defined as,
\begin{eqnarray}
\Delta \chi^2_{\rm min} = \chi^2_{\rm min}(\text{Model}) - \chi^2_{\rm min}({\rm Reference~Model}). \nonumber 
\end{eqnarray}
For the DE models with fixed neutrinos, our reference model is chosen to be the usual $\Lambda$CDM model with fixed neutrinos, while for the other set of models with free-to-vary $\sum m_{\nu,{\rm eff}}$, our reference model is $\Lambda$CDM+$\sum m_{\nu,{\rm eff}}$. If $\Delta \chi^2_{\rm min} < 0$, the model may be preferred over the reference model, and $\Delta \chi^2_{\rm min} > 0$ indicates that the reference model is favored.

Turning to the Bayesian analysis, we compute the evidence for a model $\mathcal{M}_i$ with parameter vector $\Theta$ as $\mathcal{B}_i = \int \mathcal{L}(D|\Theta,\mathcal{M}_i)\,\pi(\Theta|\mathcal{M}_i)\,d\Theta$, where $\mathcal{L}$ is the likelihood of the data $D$ and $\pi$ denotes the prior used in the analysis. 
Model comparison is performed using the Bayes factor computed for two models $i$ and $j$, $\mathcal{B}_{ij}=\mathcal{B}_i/\mathcal{B}_j$. Usually the relative log-evidence, defined as $\ln \mathcal{B}_{ij} = \ln \mathcal{B}_i - \ln \mathcal{B}_j$, is computed for this purpose. 
Here $i$ corresponds to the 
proposed model and $j$ stands for the reference model $\Lambda$CDM. Positive values of $\ln \mathcal{B}_{ij}$ indicate that the model is favored over the reference model, while negative values of $\ln \mathcal{B}_{ij}$ indicate the reverse (i.e.\ the reference model is favored over the model). Following the revised Jeffreys' scale~\cite{Kass:1995loi}, the evidence is classified as \textit{inconclusive} ($0 \leq \ln \mathcal{B}_{ij} < 1$), \textit{weak} ($1 \leq \ln \mathcal{B}_{ij} < 2.5$), \textit{moderate} ($2.5 \leq \ln \mathcal{B}_{ij} < 5$), \textit{strong} ($5 \leq \ln \mathcal{B}_{ij} < 10$), and \textit{very strong} ($\ln \mathcal{B}_{ij} \geq 10$). We use \texttt{MCEvidence}~\cite{Heavens:2017hkr,Heavens:2017afc} for computing $\ln \mathcal{B}_{ij}$.  
Note that for the DE models with fixed neutrinos, 
our reference model is chosen to be the usual $\Lambda$CDM model with fixed neutrinos, while for the other set of models with free-to-vary $\sum m_{\nu,{\rm eff}}$, our reference model is $\Lambda$CDM+$\sum m_{\nu,{\rm eff}}$. 

Note that our values for $\chi^2$ and the evidence will not agree exactly with those in, e.g.,~\cite{DESI:2025fii}, since we use different CMB data (and possibly hyperparameter settings etc.). However, the differences should not be large and the ordering of preferred models should not be affected.

\section{Results}
\label{sec-results}

We investigate the parameter constraints and goodness of fit for various dataset combinations for each of the one-parameter dark energy models.  For the six one-parameter DDE parametrizations considered in this article, the inferred neutrino mass constraints span a wide range of behaviours. While for the thawing model and the generalized thawing models, posterior distributions are shifted toward unphysical negative neutrino masses, mirage and GEDE models significantly affect the constraints  leading to $\sum m_{\nu}^{\rm eff}>0$. As we will illustrate below, these differences are mainly influenced by the nature of the dark energy equation of state. In particular, models with a lower dark energy density at high redshift shift the neutrino mass posterior toward the physical region, while crossing $w=-1$ provides a better overall fit to the observational data. In Tables~\ref{table:Model-thawing}, \ref{table:generalized-thawing-p-1}, \ref{table:generalized-thawing-p-5}, \ref{table:generalized-thawing-p-15}, \ref{table:mirage}, and \ref{table:gede} we present the cosmological constraints, and in Figs.~\ref{fig:2D}, \ref{fig:2D-all-models}, and \ref{fig:2D-gede} we show the 2D contours for the neutrino mass and the DE EoS, since our focus is to understand the interplay between dynamical dark energy and the positivity (or lack thereof) of the neutrino mass. 

\subsection{Physical Interpretation} 

Before examining the detailed results for the individual models, let us 
summarize the general results regarding neutrino mass from a 
physics perspective. 

The thawing cases, whether the calibrated 
model or the generalized ones, do not succeed in allowing a 
substantial fraction of the neutrino mass posterior to be 
positive. Thus, rapid evolution at low redshift does not 
appear to be a key factor in why $w_0w_a$CDM enables positive 
neutrino mass. 

The mirage case, despite being a one-parameter model, has all three 
properties exhibited by $w_0w_a$: recent rapid evolution, 
crossing of $w=-1$, and low dark energy density at high 
redshift. Mirage dark energy does enable positive neutrino 
mass, but with a single parameter, indicating that parameter 
space degeneracy is not a major factor. 

GEDE does not have recent rapid evolution or crossing of 
$w=-1$, but it does have low dark energy density at high 
redshift. It partly succeeds in moving the neutrino mass 
posterior toward positive values, but is limited by 
its restriction not to cross $w=-1$ (which also gives a 
poorer fit to the full data combination). When we allow 
higher $\delta$ (and hence at high redshift $w$ further 
below $-1$ and thus lower dark energy density), then 
the neutrino mass is shifted more toward positive values 
(but the overall fit becomes poorer).

\begin{figure*}
    \centering
\includegraphics[width=0.9\textwidth]{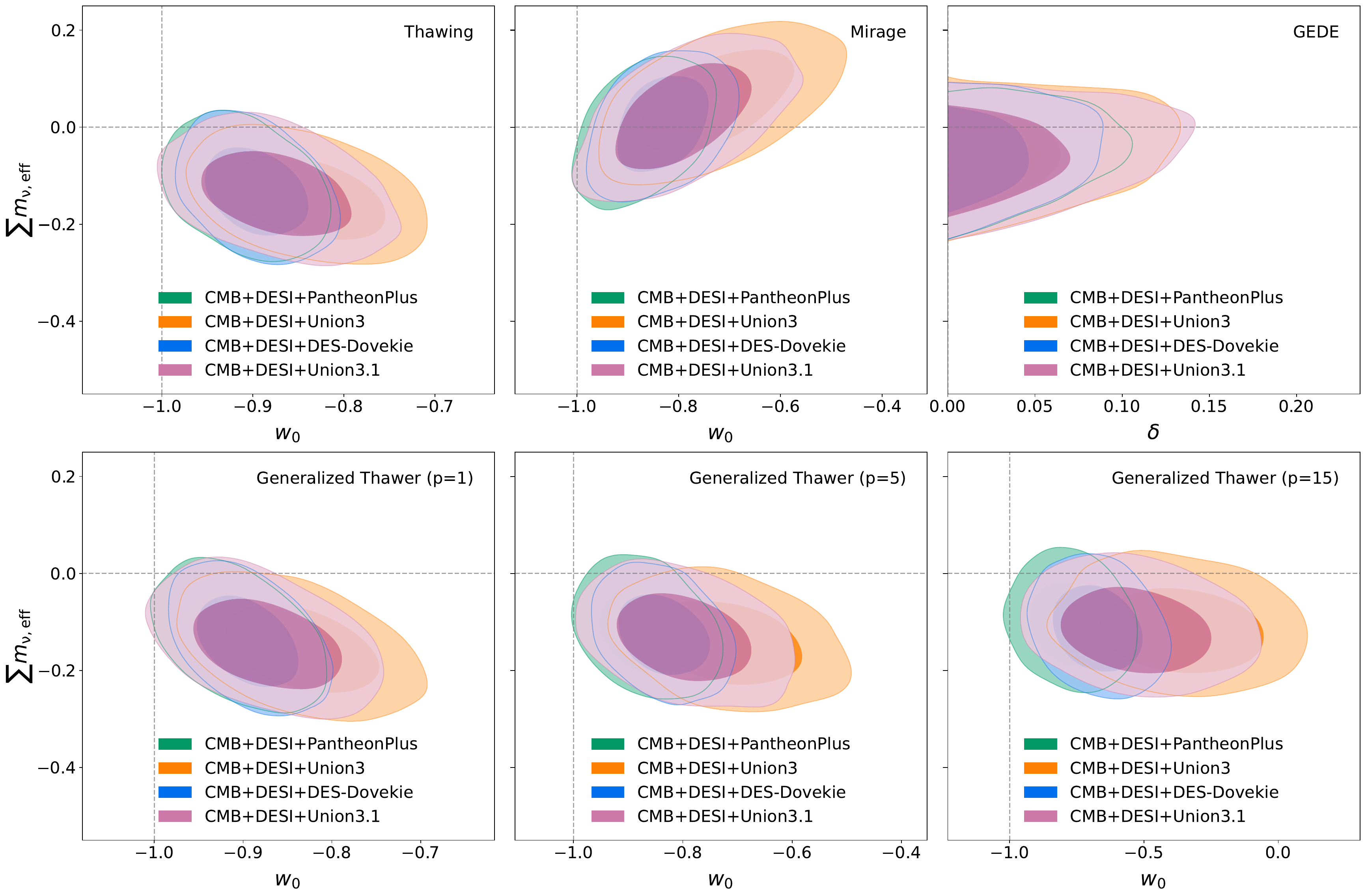}

    \caption{68\% and 95\% CL constraints on $\sum m_{\nu,{\rm eff}}$ and $w_0$ for each class of one-parameter dynamical DE models, for the combined dataset CMB+DESI+SN Ia where one of the four compilations of the SN Ia dataset is considered at a time. {\it Top row} $-$ {\bf thawing, mirage, GEDE}; {\it Bottom row} $-$ {\bf generalized thawers} with $p=1$, $5$, $15$ (from left to right).
    }
    \label{fig:2D}
\end{figure*}

\begin{figure}
    \centering
    \includegraphics[width=0.7\columnwidth]{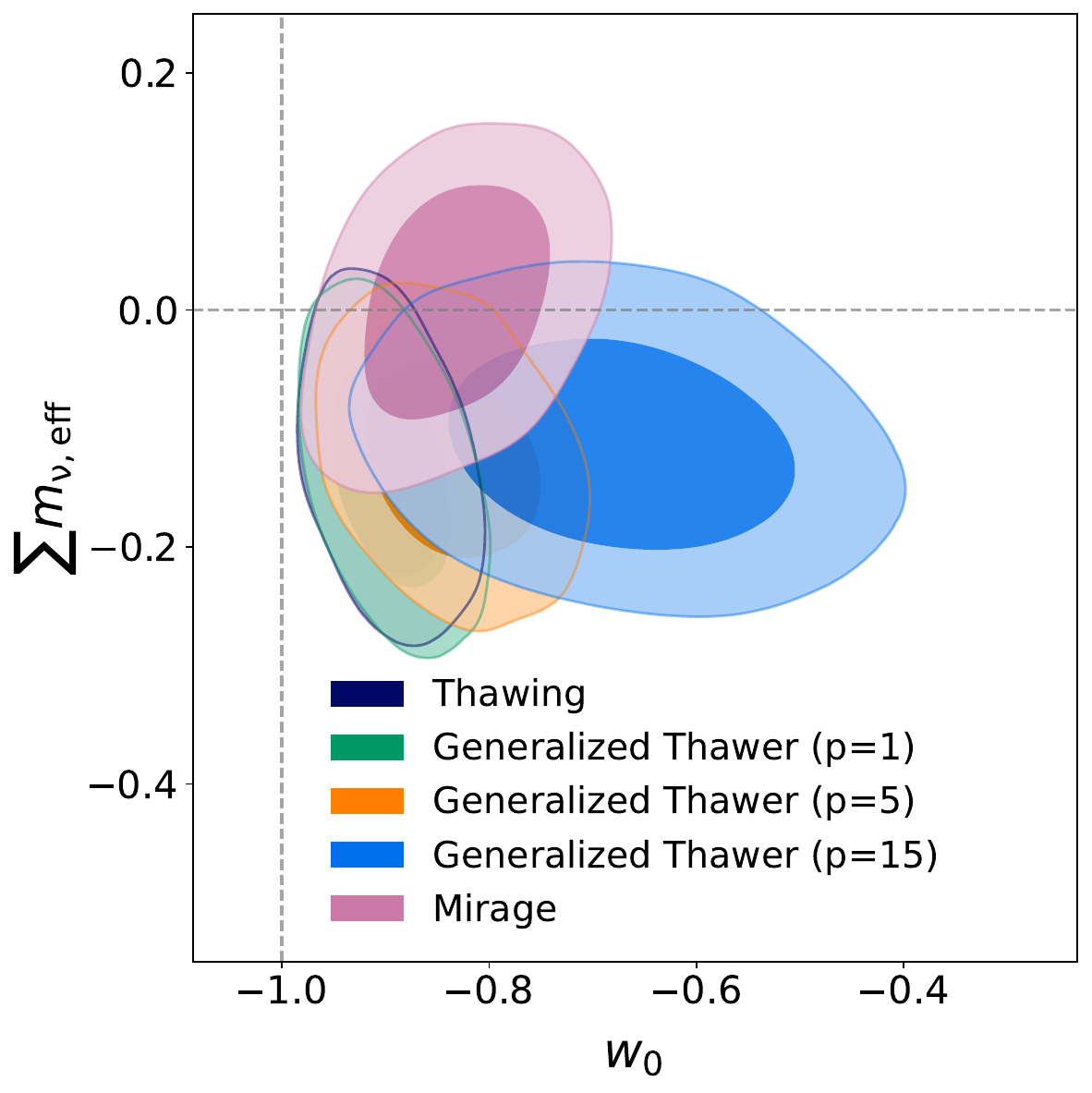}
    \caption{68\% and 95\% CL constraints on $\sum m_{\nu,{\rm eff}}$ and $w_0$ for each of the one-parameter dynamical DE models considering the combined dataset CMB+DESI+DES-Dovekie.}
    \label{fig:2D-all-models}
\end{figure}

\begin{figure}
    \centering
    \includegraphics[width=0.7\columnwidth]{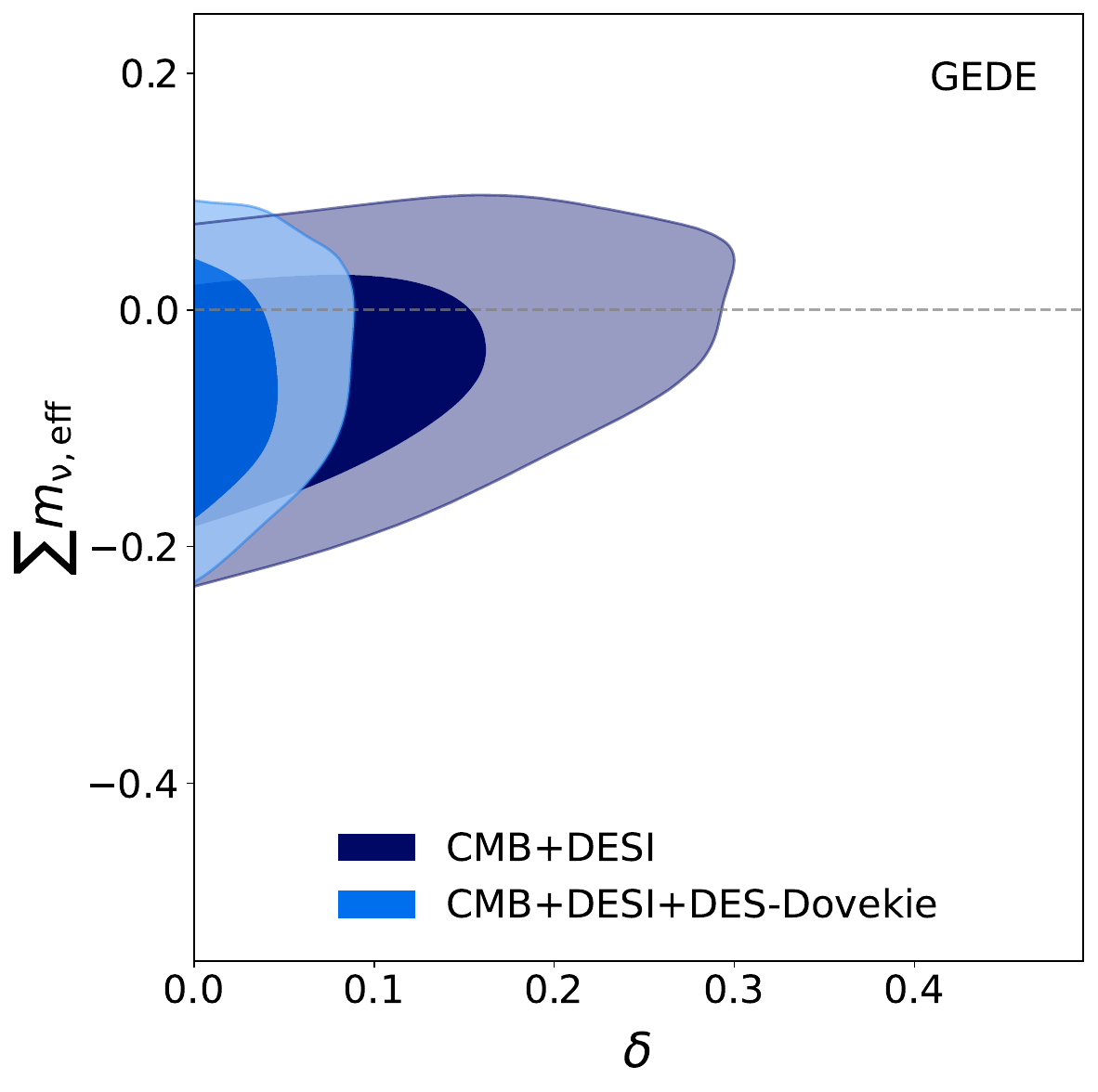}
    \caption{68\% and 95\% CL constraints on $\sum m_{\nu,{\rm eff}}$ and $\delta$ for the GEDE model considering the combined dataset CMB+DESI+DES-Dovekie, and CMB+DESI.}
    \label{fig:2D-gede}
\end{figure}

The positivity -- or lack of it -- for each 
case is summarized in Figure~\ref{fig:2D}, 
where we see that only the mirage model has 
substantial overlap with positive neutrino mass, 
while GEDE has a slight tendency in that 
direction. For easy comparison, 
Figure~\ref{fig:2D-all-models} brings together 
the five models with $w_0>-1$, showing that 
mirage dark energy is most consistent with 
positive neutrino mass. (For a clearer figure 
we show only the combination CMB+DESI+DES-Dovekie, but we saw in Figure~\ref{fig:2D} 
that all the SN datasets give similar results.) 
In Figure~\ref{fig:2D-gede} we see the GEDE 
model (plotted vs $\delta$ rather than $w_0$), 
with its modest overlap with positive neutrino 
mass. 

Thus the key elements appear to be lower dark energy density 
at high redshift to allow positive neutrino mass, and 
crossing of $w=-1$ to provide a good fit to the array of 
cosmological observations. We discuss this further in 
Section~\ref{sec-summary} after analyzing the individual 
dark energy models below.

\subsection{Thawing} 

Table~\ref{table:Model-thawing} summarizes the constraints on thawing DE at 68\% and 95\% CL. The upper half of the table corresponds to the constraints for a fixed $\sum m_{\nu,{\rm eff}}$, and the lower half of the table corresponds to the free-to-vary $\sum m_{\nu,{\rm eff}}$ case.  

In the case with fixed neutrino mass (upper half of Table~\ref{table:Model-thawing}), CMB-only data prefer a mild phantom thawer at slightly more than $1\sigma$, however, within the $2\sigma$ region $w_0 > -1$ is allowed. This preference is driven by the well-known geometrical degeneracy present in CMB-only analyses, which allows variations in the late-time expansion history to compensate for changes in other cosmological parameters. Once additional low-redshift datasets are included, this degeneracy is broken. For CMB+DESI, either phantom or normal thawers are essentially equally allowed, as is $\Lambda$. 
Once CMB and DESI are combined with any of the SN data compilations, 
the thawer lies in the quintessence regime 
at the $\sim95\%$ CL. 
According to the model comparison statistics, 
there is no strong preference for the calibrated thawer over $\Lambda$CDM. 
Although $\Delta\chi^2_{\rm min}<0$ (except for CMB+DESI in which $\Delta\chi^2_{\rm min} = 1.1$, thereby preferring the $\Lambda$CDM model), it is only mildly negative, especially considering there is one extra parameter. From Bayesian evidence analysis, we find that $\ln B_{ij} > 0$ only for CMB and CMB+DESI+Union3. However, any such preference is inconclusive or weak. 

In the case with neutrino mass free, we notice that only CMB alone prefers $w_0$ being in the phantom regime ($w_0 < -1$ at somewhat more than $1\sigma$), while 
other combined datasets favor a quintessential thawer and point to a possible deviation from $w =-1$ (CMB+DESI yields $w_0 > -1$ at $1\sigma$). In particular, this preference is stronger (at $\gtrsim 2\sigma$) when all three datasets, namely CMB, DESI, and SN Ia, are combined together. 
Concerning the total neutrino mass we find that 
$\sum m_{\nu,{\rm eff}} <0$ at $\approx2\sigma$ 
for all combined datasets; only for CMB alone is this weakened to $\approx1\sigma$. 
According to the statistical comparisons, $\Delta \chi^2_{\rm min}$ always prefers this thawing DE model to $\Lambda$CDM across all the datasets. For the Bayesian analysis, except for CMB+DESI ($\ln B_{ij} = -1.0$), all other datasets also prefer this model, if somewhat mildly.


\begingroup
\squeezetable                                   
\begin{center}  
	\begin{table*}
    \scalebox{0.85}{
		\begin{tabular}{cccccccc}
			\hline
			Parameters & CMB & CMB+DESI & CMB+DESI+PantheonPlus & CMB+DESI+Union3 & CMB+DESI+DES-Dovekie  & CMB+DESI+Union3.1 \\ \hline

            $\Omega_\mathrm{b} h^2$ & $0.02239_{-0.00015-0.00030}^{+0.00015+0.00029}$ & $0.02252_{-0.00013-0.00026}^{+0.00013+0.00026}$ & $0.02255_{-0.00012-0.00025}^{+0.00012+0.00025}$ & $0.02256_{-0.00013-0.00025}^{+0.00013+0.00025}$ & $0.02255_{-0.00013-0.00025}^{+0.00013+0.00026}$ & $0.02255_{-0.00013-0.00025}^{+0.00013+0.00025}$\\
			
			$\Omega_\mathrm{c} h^2$ & $0.1197_{-0.0012-0.0024}^{+0.0012+0.0024}$ & $0.11772_{-0.00078-0.0016}^{+0.00078+0.0015}$ & $0.11742_{-0.00071-0.0013}^{+0.00069+0.0014}$ & $0.11725_{-0.00072-0.0014}^{+0.00072+0.0014}$ & $0.11735_{-0.00070-0.0013}^{+0.00068+0.0014}$ & $0.11737_{-0.00071-0.0014}^{+0.00071+0.0014}$\\
			
			$100\theta_\mathrm{MC}$ & $1.04187_{-0.00030-0.00060}^{+0.00029+0.00058}$ & $1.04207_{-0.00027-0.00055}^{+0.00028+0.00053}$ & $1.04211_{-0.00029-0.00055}^{+0.00027+0.00056}$ & $1.04211_{-0.00028-0.00056}^{+0.00028+0.00054}$ & $1.04211_{-0.00027-0.00055}^{+0.00027+0.00053}$ & $1.04212_{-0.00028-0.00056}^{+0.00027+0.00053}$\\
			
			$\tau$ & $0.0533_{-0.0073-0.014}^{+0.0071+0.015}$ & $0.0604_{-0.0079-0.015}^{+0.0067+0.015}$ & $0.0615_{-0.0079-0.015}^{+0.0072+0.016}$ & $0.0622_{-0.0081-0.014}^{+0.0069+0.016}$ & $0.0617_{-0.0076-0.014}^{+0.0068+0.015}$ & $0.0616_{-0.0079-0.015}^{+0.0069+0.016}$\\
			
			$n_\mathrm{s}$ & $0.9652_{-0.0042-0.0084}^{+0.0042+0.0082}$ & $0.9701_{-0.0036-0.0070}^{+0.0036+0.0069}$ & $0.9710_{-0.0033-0.0067}^{+0.0033+0.0067}$ & $0.9714_{-0.0034-0.0066}^{+0.0034+0.0065}$ & $0.9711_{-0.0033-0.0067}^{+0.0035+0.0069}$ & $0.9711_{-0.0034-0.0067}^{+0.0033+0.0069}$\\
			
			$\ln(10^{10} A_\mathrm{s})$ & $3.041_{-0.014-0.028}^{+0.014+0.028}$ & $3.052_{-0.016-0.028}^{+0.014+0.031}$ & $3.054_{-0.015-0.030}^{+0.015+0.031}$ & $3.055_{-0.016-0.029}^{+0.014+0.031}$ & $3.054_{-0.014-0.029}^{+0.014+0.029}$ & $3.054_{-0.016-0.029}^{+0.014+0.031}$\\
			
			$w_0$ & $-1.90_{-0.78-0.85}^{+0.27+1.2}$ & $-1.012_{-0.089-0.18}^{+0.088+0.17}$ & $-0.941_{-0.037-0.071}^{+0.036+0.074}$ & $-0.904_{-0.052-0.10}^{+0.053+0.11}$ & $-0.933_{-0.035-0.067}^{+0.034+0.069}$ & $-0.932_{-0.048-0.094}^{+0.047+0.095}$\\
			
			$\Omega_\mathrm{m}$ & $0.203_{-0.082-0.090}^{+0.020+0.16}$ & $0.300_{-0.012-0.023}^{+0.012+0.023}$ & $0.3087_{-0.0057-0.011}^{+0.0052+0.011}$ & $0.3138_{-0.0074-0.014}^{+0.0075+0.015}$ & $0.3096_{-0.0050-0.010}^{+0.0050+0.010}$ & $0.3099_{-0.0067-0.013}^{+0.0068+0.013}$\\
			
			$\sigma_8$ & $0.964_{-0.058-0.20}^{+0.137+0.15}$ & $0.809_{-0.015-0.029}^{+0.016+0.031}$ & $0.798_{-0.009-0.017}^{+0.009+0.017}$ & $0.793_{-0.010-0.020}^{+0.010+0.020}$ & $0.797_{-0.008-0.016}^{+0.008+0.016}$ & $0.797_{-0.010-0.019}^{+0.010+0.019}$\\
			
			$H_0$ [km/s/Mpc] & $87_{-9-26}^{+18+22}$ & $68.6_{-1.5-2.7}^{+1.3+2.9}$ & $67.50_{-0.58-1.2}^{+0.58+1.1}$ & $66.92_{-0.81-1.6}^{+0.79+1.6}$ & $67.38_{-0.52-1.0}^{+0.53+1.0}$ & $67.36_{-0.74-1.5}^{+0.74+1.5}$\\
			
			$S_8$ & $0.767_{-0.054-0.071}^{+0.031+0.087}$ & $0.808_{-0.008-0.017}^{+0.008+0.017}$ & $0.810_{-0.008-0.016}^{+0.008+0.017}$ & $0.811_{-0.009-0.016}^{+0.008+0.017}$ & $0.809_{-0.008-0.015}^{+0.008+0.016}$ & $0.810_{-0.009-0.016}^{+0.009+0.017}$\\
			
			$r_{\rm{drag}}$ [Mpc] & $147.16_{-0.27-0.51}^{+0.27+0.54}$ & $147.53_{-0.21-0.41}^{+0.21+0.42}$ & $147.58_{-0.20-0.40}^{+0.20+0.39}$ & $147.62_{-0.20-0.39}^{+0.20+0.39}$ & $147.60_{-0.21-0.40}^{+0.20+0.40}$ & $147.60_{-0.20-0.40}^{+0.20+0.40}$\\
			
			\hline
			
			$\Delta\chi^2_{\rm min}$ & $-1.2$ & $1.1$ & $-1.5$ & $-2.5$ & $-3.0$ & $-0.9$  \\
			
			$\ln \mathcal{B}_{ij}$ & $1.2$ & $-1.9$ & $-1.6$ & $0.3$ & $-0.6$ & $-1.5$\\

		\end{tabular}}
        \scalebox{0.85}{
        \begin{tabular}{cccccccc}
			\hline
			Parameters & CMB & CMB+DESI & CMB+DESI+PantheonPlus & CMB+DESI+Union3 & CMB+DESI+DES-Dovekie & CMB+DESI+Union3.1  \\ \hline

            $\Omega_\mathrm{b} h^2$ & $0.02247_{-0.00017-0.00034}^{+0.00017+0.00036}$ & $0.02245_{-0.00013-0.00026}^{+0.00014+0.00026}$ & $0.02245_{-0.00013-0.00025}^{+0.00013+0.00025}$ & $0.02247_{-0.00013-0.00024}^{+0.00013+0.00026}$ & $0.02245_{-0.00013-0.00025}^{+0.00013+0.00026}$ & $0.02245_{-0.00013-0.00026}^{+0.00013+0.00025}$\\
			
			$\Omega_\mathrm{c} h^2$ & $0.1188_{-0.0015-0.0030}^{+0.0015+0.0031}$ & $0.11902_{-0.00090-0.0017}^{+0.00090+0.0017}$ & $0.11897_{-0.00089-0.0017}^{+0.00090+0.0017}$ & $0.11885_{-0.00084-0.0017}^{+0.00087+0.0017}$ & $0.11896_{-0.00089-0.0017}^{+0.00089+0.0018}$ & $0.11890_{-0.00090-0.0018}^{+0.00089+0.0017}$\\
			
			$100\theta_\mathrm{MC}$ & $1.04193_{-0.00030-0.00060}^{+0.00030+0.00060}$ & $1.04189_{-0.00028-0.00061}^{+0.00029+0.00056}$ & $1.04191_{-0.00028-0.00055}^{+0.00028+0.00057}$ & $1.04191_{-0.00028-0.00054}^{+0.00028+0.00056}$ & $1.04189_{-0.00029-0.00056}^{+0.00029+0.00055}$ & $1.04191_{-0.00028-0.00054}^{+0.00028+0.00055}$\\
			
			$\tau$ & $0.0512_{-0.0077-0.017}^{+0.0078+0.016}$ & $0.0511_{-0.0076-0.016}^{+0.0073+0.016}$ & $0.0518_{-0.0074-0.015}^{+0.0075+0.015}$ & $0.0515_{-0.0075-0.016}^{+0.0076+0.016}$ & $0.0515_{-0.0076-0.015}^{+0.0076+0.016}$ & $0.0515_{-0.0077-0.015}^{+0.0075+0.015}$\\
			
			$n_\mathrm{s}$ & $0.9678_{-0.0048-0.0099}^{+0.0050+0.0099}$ & $0.9673_{-0.0037-0.0072}^{+0.0037+0.0071}$ & $0.9674_{-0.0037-0.0069}^{+0.0035+0.0072}$ & $0.9675_{-0.0037-0.0068}^{+0.0034+0.0071}$ & $0.9672_{-0.0035-0.0071}^{+0.0036+0.0071}$ & $0.9674_{-0.0036-0.0072}^{+0.0036+0.0071}$\\
			
			$\ln(10^{10} A_\mathrm{s})$ & $3.035_{-0.015-0.034}^{+0.017+0.031}$ & $3.035_{-0.015-0.031}^{+0.015+0.030}$ & $3.036_{-0.015-0.029}^{+0.015+0.029}$ & $3.036_{-0.015-0.031}^{+0.015+0.030}$ & $3.036_{-0.015-0.030}^{+0.015+0.031}$ & $3.035_{-0.015-0.030}^{+0.015+0.030}$\\
			
\rowcolor{LightSkyBlue} 
        $\sum m_{\nu,{\rm eff}}$ & $-0.12_{-0.16-0.35}^{+0.17+0.34}$ & $-0.132_{-0.064-0.13}^{+0.063+0.14}$ & $-0.127_{-0.063-0.12}^{+0.056+0.13}$ & $-0.144_{-0.061-0.11}^{+0.054+0.12}$ & $-0.131_{-0.063-0.12}^{+0.057+0.14}$ & $-0.135_{-0.059-0.12}^{+0.057+0.13}$\\
			
			$w_0$ & $-1.78_{-0.90-0.97}^{+0.32+1.3}$ & $-0.90_{-0.10-0.19}^{+0.10+0.19}$ & $-0.906_{-0.037-0.077}^{+0.038+0.072}$ & $-0.840_{-0.055-0.11}^{+0.056+0.11}$ & $-0.897_{-0.036-0.070}^{+0.036+0.072}$ & $-0.874_{-0.051-0.10}^{+0.053+0.11}$\\
			
			$\Omega_\mathrm{m}$ & $0.200_{-0.097-0.12}^{+0.031+0.18}$ & $0.306_{-0.012-0.024}^{+0.012+0.024}$ & $0.3047_{-0.0055-0.011}^{+0.0055+0.011}$ & $0.3127_{-0.0073-0.014}^{+0.0072+0.015}$ & $0.3057_{-0.0054-0.011}^{+0.0056+0.011}$ & $0.3086_{-0.0071-0.014}^{+0.0071+0.014}$\\
			
			$\sigma_8$ & $0.99_{-0.11-0.24}^{+0.14+0.22}$ & $0.827_{-0.016-0.032}^{+0.016+0.033}$ & $0.829_{-0.013-0.026}^{+0.013+0.026}$ & $0.821_{-0.013-0.026}^{+0.013+0.026}$ & $0.828_{-0.013-0.027}^{+0.013+0.026}$ & $0.824_{-0.013-0.028}^{+0.013+0.027}$\\
			
			$H_0$ [km/s/Mpc] & $88_{-15-30}^{+18+27}$ & $67.6_{-1.5-2.8}^{+1.4+2.9}$ & $67.72_{-0.56-1.1}^{+0.56+1.1}$ & $66.77_{-0.80-1.5}^{+0.79+1.6}$ & $67.59_{-0.57-1.1}^{+0.56+1.1}$ & $67.25_{-0.79-1.5}^{+0.77+1.6}$\\
			
			$S_8$ & $0.768_{-0.062-0.088}^{+0.042+0.10}$ & $0.836_{-0.013-0.025}^{+0.012+0.025}$ & $0.835_{-0.011-0.022}^{+0.011+0.023}$ & $0.838_{-0.012-0.022}^{+0.011+0.022}$ & $0.835_{-0.012-0.024}^{+0.012+0.023}$ & $0.836_{-0.012-0.024}^{+0.012+0.023}$\\
			
			$r_{\rm{drag}}$ [Mpc] & $147.34_{-0.33-0.66}^{+0.33+0.65}$ & $147.29_{-0.22-0.43}^{+0.22+0.43}$ & $147.30_{-0.23-0.44}^{+0.22+0.45}$ & $147.31_{-0.22-0.44}^{+0.22+0.42}$ & $147.30_{-0.22-0.45}^{+0.22+0.44}$ & $147.32_{-0.22-0.44}^{+0.23+0.44}$\\
			
			\hline
			
			$\Delta\chi^2_{\rm min}$ & $-1.2$ & $-0.3$ & $-5.0$ & $-7.3$ & $-7.2$ & $-4.2$ \\
			
			$\ln \mathcal{B}_{ij}$ & $1.3$ & $-1.0$ & $0.4$ & $2.1$ & $1.2$ & $0.7$ \\			
			
			\hline                                                        
		\end{tabular}}                                                       
		\caption{Summary of the observational constraints on various free and derived cosmological parameters of {\bf thawing dark energy}, 
        at 68\% and 95\% CL using several observational datasets. The upper half of the table corresponds to the scenario with neutrino mass fixed, while the lower half of the table corresponds to the scenario including neutrino mass as a free parameter.  
        }
		\label{table:Model-thawing}                   
	\end{table*}                                     
\end{center}
\endgroup

\begingroup
\squeezetable                                   
\begin{center}  
	\begin{table*}
    \scalebox{0.85}{
		\begin{tabular}{cccccccc}
			\hline
			Parameters & CMB & CMB+DESI & CMB+DESI+PantheonPlus & CMB+DESI+Union3 & CMB+DESI+DES-Dovekie &  CMB+DESI+Union3.1 \\ \hline

            $\Omega_\mathrm{b} h^2$ & $0.02241_{-0.00015-0.00029}^{+0.00015+0.00030}$ & $0.02252_{-0.00013-0.00026}^{+0.00013+0.00026}$ & $0.02255_{-0.00013-0.00026}^{+0.00013+0.00026}$ & $0.02256_{-0.00013-0.00025}^{+0.00013+0.00025}$ & $0.02255_{-0.00013-0.00026}^{+0.00013+0.00026}$ & $0.02255_{-0.00013-0.00025}^{+0.00013+0.00025}$ \\
				
				$\Omega_\mathrm{c} h^2$ & $0.1194_{-0.0012-0.0024}^{+0.0012+0.0025}$ & $0.11780_{-0.00080-0.0016}^{+0.00080+0.0016}$ & $0.11742_{-0.00075-0.0015}^{+0.00075+0.0014}$ & $0.11724_{-0.00073-0.0014}^{+0.00073+0.0015}$ & $0.11733_{-0.00069-0.0014}^{+0.00071+0.0014}$ & $0.11736_{-0.00071-0.0014}^{+0.00070+0.0014}$ \\
				
				$100\theta_\mathrm{MC}$ & $1.04189_{-0.00029-0.00058}^{+0.00030+0.00059}$ & $1.04206_{-0.00028-0.00056}^{+0.00028+0.00054}$ & $1.04212_{-0.00028-0.00056}^{+0.00028+0.00055}$ & $1.04213_{-0.00026-0.00052}^{+0.00026+0.00052}$ & $1.04211_{-0.00027-0.00052}^{+0.00027+0.00054}$ & $1.04212_{-0.00027-0.00054}^{+0.00027+0.00056}$ \\
				
				$\tau$ & $0.0528_{-0.0071-0.014}^{+0.0072+0.014}$ & $0.0600_{-0.0076-0.014}^{+0.0068+0.015}$ & $0.0612_{-0.0082-0.014}^{+0.0071+0.016}$ & $0.0623_{-0.0078-0.014}^{+0.0071+0.015}$ & $0.0620_{-0.0077-0.014}^{+0.0069+0.015}$ & $0.0617_{-0.0079-0.014}^{+0.0069+0.016}$ \\
				
				$n_\mathrm{s}$ & $0.9660_{-0.0042-0.0082}^{+0.0042+0.0083}$ & $0.9700_{-0.0035-0.0071}^{+0.0035+0.0068}$ & $0.9709_{-0.0035-0.0065}^{+0.0034+0.0068}$ & $0.9713_{-0.0036-0.0071}^{+0.0036+0.0069}$ & $0.9712_{-0.0035-0.0064}^{+0.0034+0.0068}$ & $0.9710_{-0.0034-0.0066}^{+0.0034+0.0068}$ \\
				
				$\ln(10^{10} A_\mathrm{s})$ & $3.040_{-0.014-0.028}^{+0.014+0.028}$ & $3.052_{-0.015-0.028}^{+0.014+0.030}$ & $3.053_{-0.016-0.029}^{+0.014+0.031}$ & $3.055_{-0.015-0.029}^{+0.015+0.030}$ & $3.055_{-0.014-0.028}^{+0.014+0.030}$ & $3.054_{-0.015-0.028}^{+0.015+0.031}$ \\
				
				$w_0$ & $<-1.973<-1.021$ & $-1.030_{-0.085-0.18}^{+0.093+0.169}$ & $-0.943_{-0.038-0.075}^{+0.038+0.076}$ & $-0.907_{-0.051-0.10}^{+0.051+0.10}$ & $-0.935_{-0.035-0.071}^{+0.036+0.069}$ & $-0.936_{-0.047-0.097}^{+0.049+0.096}$ \\
				
			$\Omega_\mathrm{m}$ & $0.173_{-0.073-0.079}^{+0.016+0.141}$ & $0.297_{-0.012-0.023}^{+0.012+0.022}$ & $0.3084_{-0.0056-0.011}^{+0.0056+0.011}$ & $0.3131_{-0.0077-0.014}^{+0.0070+0.015}$ & $0.3093_{-0.0051-0.0098}^{+0.0050+0.011}$ & $0.3092_{-0.0069-0.013}^{+0.0068+0.014}$ \\
            
		$\sigma_8$ & $1.019_{-0.059-0.22}^{+0.15+0.17}$ & $0.812_{-0.017-0.029}^{+0.015+0.032}$ & $0.798_{-0.009-0.018}^{+0.009+0.017}$ & $0.793_{-0.010-0.019}^{+0.010+0.020}$ & $0.797_{-0.008-0.016}^{+0.008+0.016}$ & $0.797_{-0.010-0.019}^{+0.010+0.019}$ \\
				
				$H_0$ [km/s/Mpc] & $95_{-10-29}^{+21+24}$ & $68.9_{-1.5-2.6}^{+1.3+2.9}$ & $67.54_{-0.59-1.2}^{+0.59+1.2}$ & $66.99_{-0.78-1.6}^{+0.81+1.5}$ & $67.41_{-0.55-1.1}^{+0.54+1.1}$ & $67.43_{-0.75-1.5}^{+0.76+1.5}$ \\
				
				$S_8$ & $0.745_{-0.059-0.075}^{+0.031+0.092}$ & $0.808_{-0.008-0.016}^{+0.008+0.016}$ & $0.809_{-0.009-0.017}^{+0.009+0.017}$ & $0.810_{-0.009-0.017}^{+0.008+0.017}$ & $0.809_{-0.008-0.016}^{+0.008+0.016}$ & $0.809_{-0.008-0.017}^{+0.008+0.017}$ \\
				
				$r_{\rm{drag}}$ [Mpc] & $147.22_{-0.26-0.54}^{+0.27+0.53}$ & $147.52_{-0.21-0.42}^{+0.22+0.42}$ & $147.58_{-0.20-0.41}^{+0.20+0.40}$ & $147.62_{-0.19-0.42}^{+0.21+0.40}$ & $147.61_{-0.20-0.39}^{+0.20+0.39}$ & $147.60_{-0.20-0.40}^{+0.21+0.39}$ \\
				
				\hline
				
				$\Delta\chi^2_{\rm min}$ & $-1.9$ & $1.1$ & $-1.0$ & $-2.1$ & $-2.8$ & $-0.6$ \\
				
				$\ln \mathcal{B}_{ij}$ & $1.5$ & $-2.1$ & $-1.7$ & $-0.1$ & $-0.6$ & $-1.5$ \\
			                   
		\end{tabular}}   
		\scalebox{0.85}{
		\begin{tabular}{cccccccc}
			\hline
			Parameters & CMB & CMB+DESI & CMB+DESI+PantheonPlus & CMB+DESI+Union3 & CMB+DESI+DES-Dovekie &  CMB+DESI+Union3.1 \\ \hline
			
			$\Omega_\mathrm{b} h^2$ & $0.02248_{-0.00018-0.00035}^{+0.00018+0.00036}$ & $0.02246_{-0.00013-0.00026}^{+0.00013+0.00026}$ & $0.02245_{-0.00013-0.00026}^{+0.00013+0.00025}$ & $0.02247_{-0.00013-0.00025}^{+0.00013+0.00025}$ & $0.02246_{-0.00014-0.00027}^{+0.00013+0.00027}$ & $0.02247_{-0.00013-0.00025}^{+0.00013+0.00025}$\\
				
				$\Omega_\mathrm{c} h^2$ & $0.1187_{-0.0015-0.0030}^{+0.0015+0.0030}$ & $0.11900_{-0.00087-0.0018}^{+0.00087+0.0017}$ & $0.11898_{-0.00087-0.0017}^{+0.00086+0.0017}$ & $0.11884_{-0.00087-0.0018}^{+0.00088+0.0018}$ & $0.11897_{-0.00090-0.0017}^{+0.00089+0.0018}$ & $0.11892_{-0.00090-0.0017}^{+0.00088+0.0018}$ \\
				
				$100\theta_\mathrm{MC}$ & $1.04195_{-0.00030-0.00058}^{+0.00030+0.00059}$ & $1.04190_{-0.00028-0.00054}^{+0.00027+0.00055}$ & $1.04190_{-0.00028-0.00054}^{+0.00028+0.00055}$ & $1.04191_{-0.00028-0.00055}^{+0.00028+0.00055}$ & $1.04190_{-0.00028-0.00055}^{+0.00028+0.00058}$ & $1.04190_{-0.00030-0.00055}^{+0.00028+0.00057}$\\
				
				$\tau$ & $0.0504_{-0.0078-0.016}^{+0.0078+0.016}$ & $0.0515_{-0.0076-0.015}^{+0.0076+0.016}$ & $0.0515_{-0.0074-0.016}^{+0.0074+0.015}$ & $0.0513_{-0.0076-0.017}^{+0.0076+0.016}$ & $0.0511_{-0.0077-0.015}^{+0.0075+0.016}$ & $0.0513_{-0.0079-0.017}^{+0.0081+0.016}$ \\
				
				$n_\mathrm{s}$ & $0.9678_{-0.0049-0.0098}^{+0.0049+0.0097}$ & $0.9673_{-0.0035-0.0068}^{+0.0036+0.0071}$ & $0.9673_{-0.0036-0.0069}^{+0.0035+0.0070}$ & $0.9676_{-0.0037-0.0072}^{+0.0037+0.0071}$ & $0.9673_{-0.0037-0.0074}^{+0.0037+0.0071}$ & $0.9674_{-0.0037-0.0074}^{+0.0037+0.0073}$\\
				
				$\ln(10^{10} A_\mathrm{s})$ & $3.033_{-0.016-0.034}^{+0.016+0.033}$ & $3.036_{-0.015-0.030}^{+0.015+0.031}$ & $3.036_{-0.015-0.031}^{+0.015+0.030}$ & $3.035_{-0.015-0.032}^{+0.015+0.031}$ & $3.035_{-0.015-0.030}^{+0.015+0.030}$ & $3.035_{-0.015-0.033}^{+0.016+0.032}$
                \\
				
\rowcolor{LightSkyBlue} 
$\sum m_{\nu, {\rm eff}}$ & $-0.09_{-0.17-0.36}^{+0.17+0.33}$ & $-0.137_{-0.066-0.13}^{+0.064+0.13}$ & $-0.133_{-0.061-0.12}^{+0.060+0.13}$ & $-0.154_{-0.060-0.12}^{+0.059+0.13}$ & $-0.138_{-0.063-0.12}^{+0.061+0.13}$ & $-0.144_{-0.066-0.12}^{+0.056+0.14}$ \\
				
				$w_0$ & $<-1.720<-0.661$ & $-0.894_{-0.096-0.19}^{+0.099+0.19}$ & $-0.903_{-0.040-0.081}^{+0.040+0.077}$ & $-0.836_{-0.056-0.11}^{+0.056+0.11}$ & $-0.895_{-0.037-0.072}^{+0.037+0.075}$ & $-0.874_{-0.054-0.11}^{+0.054+0.11}$\\
				
		$\Omega_\mathrm{m}$ & $0.178_{-0.091-0.11}^{+0.028+0.17}$ & $0.306_{-0.012-0.022}^{+0.012+0.023}$ & $0.3046_{-0.0055-0.011}^{+0.0054+0.011}$ & $0.3124_{-0.0072-0.014}^{+0.0072+0.014}$ & $0.3053_{-0.0052-0.011}^{+0.0053+0.011}$ & $0.3078_{-0.0071-0.014}^{+0.0072+0.014}$\\
				
		$\sigma_8$ & $1.03_{-0.12-0.26}^{+0.16+0.23}$ & $0.828_{-0.016-0.031}^{+0.016+0.031}$ & $0.829_{-0.013-0.025}^{+0.013+0.025}$ & $0.822_{-0.013-0.025}^{+0.013+0.026}$ & $0.828_{-0.013-0.026}^{+0.013+0.026}$ & $0.826_{-0.014-0.027}^{+0.014+0.026}$\\
				
				$H_0$ [km/s/Mpc] & $94_{-17-34}^{+21+31}$ & $67.6_{-1.4-2.7}^{+1.4+2.8}$ & $67.70_{-0.59-1.1}^{+0.56+1.2}$ & $66.76_{-0.79-1.5}^{+0.78+1.6}$ & $67.61_{-0.54-1.1}^{+0.54+1.1}$ & $67.31_{-0.77-1.5}^{+0.79+1.5}$\\
				
				$S_8$ & $0.751_{-0.068-0.092}^{+0.043+0.11}$ & $0.836_{-0.012-0.025}^{+0.013+0.024}$ & $0.835_{-0.012-0.022}^{+0.011+0.023}$ & $0.838_{-0.012-0.023}^{+0.012+0.023}$ & $0.835_{-0.012-0.023}^{+0.012+0.023}$ & $0.836_{-0.011-0.024}^{+0.012+0.023}$\\
				
				$r_{\rm{drag}}$ [Mpc] & $147.35_{-0.32-0.63}^{+0.32+0.63}$ & $147.29_{-0.22-0.42}^{+0.22+0.44}$ & $147.30_{-0.22-0.43}^{+0.22+0.42}$ & $147.32_{-0.22-0.42}^{+0.22+0.45}$ & $147.30_{-0.22-0.45}^{+0.22+0.42}$ & $147.30_{-0.22-0.43}^{+0.22+0.42}$\\
				
				\hline
				
				$\Delta\chi^2_{\rm min}$ & $-1.6$ & $-0.4$ & $-4.8$ & $-7.5$ & $-6.8$ & $-4.3$ \\
				
				$\ln \mathcal{B}_{ij}$ & $1.5$ & $-1.0$ & $0.5$ & $2.0$ & $1.2$ & $0.7$ \\			
			
			\hline                                                        
		\end{tabular}}                                                       
		\caption{Summary of the observational constraints on various free and derived cosmological parameters of 
the $p=1$ {\bf generalized thawing dark energy}, 
at 68\% and 95\% CL using several observational datasets. Note that for the CMB-alone case we quote the upper limits on $w_0$ at 68\% and 95\% CL. 
The upper half of the table corresponds to the scenario with neutrino mass fixed, while the lower half of the table corresponds to the scenario including neutrino mass as a free parameter.}
		\label{table:generalized-thawing-p-1}                   
	\end{table*}                                     
\end{center}
\endgroup

\begingroup
\squeezetable                                   
\begin{center}  
	\begin{table*}
    \scalebox{0.85}{
		\begin{tabular}{cccccccc}
			\hline
			Parameters & CMB & CMB+DESI & CMB+DESI+PantheonPlus & CMB+DESI+Union3 & CMB+DESI+DES-Dovekie  & CMB+DESI+Union3.1 \\ \hline

          	$\Omega_\mathrm{b} h^2$ & $0.02238_{-0.00015-0.00028}^{+0.00014+0.00028}$ & $0.02252_{-0.00013-0.00025}^{+0.00013+0.00025}$ & $0.02255_{-0.00013-0.00026}^{+0.00012+0.00026}$ & $0.02256_{-0.00012-0.00025}^{+0.00012+0.00025}$ & $0.02255_{-0.00013-0.00026}^{+0.00013+0.00025}$ & $0.02256_{-0.00012-0.00026}^{+0.00012+0.00024}$ \\
				
				$\Omega_\mathrm{c} h^2$ & $0.1197_{-0.0012-0.0024}^{+0.0012+0.0023}$ & $0.11774_{-0.00079-0.0016}^{+0.00078+0.0015}$ & $0.11743_{-0.00067-0.0013}^{+0.00067+0.0014}$ & $0.11725_{-0.00069-0.0014}^{+0.00068+0.0014}$ & $0.11736_{-0.00069-0.0013}^{+0.00069+0.0013}$ & $0.11737_{-0.00068-0.0013}^{+0.00067+0.0014}$ \\
				
				$100\theta_\mathrm{MC}$ & $1.04186_{-0.00029-0.00058}^{+0.00029+0.00060}$ & $1.04209_{-0.00028-0.00055}^{+0.00027+0.00053}$ & $1.04211_{-0.00028-0.00054}^{+0.00027+0.00053}$ & $1.04213_{-0.00027-0.00055}^{+0.00027+0.00055}$ & $1.04210_{-0.00028-0.00055}^{+0.00029+0.00053}$ & $1.04210_{-0.00028-0.00054}^{+0.00028+0.00054}$ \\
				
				$\tau$ & $0.0535_{-0.0074-0.015}^{+0.0074+0.015}$ & $0.0606_{-0.0078-0.014}^{+0.0067+0.015}$ & $0.0615_{-0.0082-0.014}^{+0.0070+0.016}$ & $0.0621_{-0.0081-0.014}^{+0.0070+0.016}$ & $0.0618_{-0.0080-0.015}^{+0.0071+0.016}$ & $0.0619_{-0.0078-0.014}^{+0.0069+0.016}$ \\
				
				$n_\mathrm{s}$ & $0.9651_{-0.0040-0.0078}^{+0.0040+0.0079}$ & $0.9701_{-0.0035-0.0069}^{+0.0035+0.0069}$ & $0.9709_{-0.0032-0.0066}^{+0.0033+0.0065}$ & $0.9714_{-0.0034-0.0068}^{+0.0035+0.0067}$ & $0.9710_{-0.0034-0.0066}^{+0.0034+0.0065}$ & $0.9710_{-0.0032-0.0067}^{+0.0034+0.0066}$ \\
				
				$\ln(10^{10} A_\mathrm{s})$ & $3.042_{-0.014-0.028}^{+0.014+0.029}$ & $3.053_{-0.016-0.027}^{+0.014+0.030}$ & $3.054_{-0.016-0.028}^{+0.014+0.032}$ & $3.055_{-0.016-0.028}^{+0.014+0.031}$ & $3.054_{-0.016-0.030}^{+0.015+0.031}$ & $3.055_{-0.016-0.028}^{+0.014+0.032}$ \\
				
				$w_0$ & $<-1.66<-0.21$ & $-1.02_{-0.17-0.35}^{+0.17+0.34}$ & $-0.907_{-0.055-0.11}^{+0.053+0.11}$ & $-0.810_{-0.086-0.16}^{+0.080+0.16}$ & $-0.886_{-0.053-0.11}^{+0.052+0.11}$ & $-0.870_{-0.080-0.15}^{+0.079+0.16}$ \\
				
		$\Omega_\mathrm{m}$ & $0.233_{-0.087-0.094}^{+0.021+0.19}$ & $0.299_{-0.016-0.031}^{+0.016+0.032}$ & $0.3093_{-0.0058-0.011}^{+0.0057+0.012}$ & $0.3186_{-0.0082-0.016}^{+0.0083+0.016}$ & $0.3112_{-0.0057-0.011}^{+0.0057+0.011}$ & $0.3129_{-0.0078-0.015}^{+0.0077+0.016}$ \\
				
		$\sigma_8$ & $0.917_{-0.049-0.18}^{+0.12+0.14}$ & $0.810_{-0.019-0.036}^{+0.019+0.040}$ & $0.798_{-0.009-0.016}^{+0.008+0.017}$ & $0.788_{-0.010-0.020}^{+0.010+0.020}$ & $0.796_{-0.008-0.016}^{+0.008+0.016}$ & $0.794_{-0.010-0.020}^{+0.010+0.020}$ \\
				
				$H_0$ [km/s/Mpc] & $81_{-7-23}^{+16+18}$ & $68.7_{-2.0-3.6}^{+1.8+4.0}$ & $67.44_{-0.59-1.2}^{+0.60+1.2}$ & $66.42_{-0.88-1.7}^{+0.86+1.7}$ & $67.22_{-0.60-1.1}^{+0.56+1.1}$ & $67.04_{-0.84-1.7}^{+0.84+1.6}$ \\
				
				$S_8$ & $0.785_{-0.050-0.064}^{+0.027+0.081}$ & $0.808_{-0.009-0.016}^{+0.009+0.017}$ & $0.810_{-0.008-0.016}^{+0.008+0.017}$ & $0.812_{-0.008-0.016}^{+0.008+0.017}$ & $0.810_{-0.008-0.017}^{+0.008+0.016}$ & $0.811_{-0.008-0.017}^{+0.008+0.017}$ \\
				
				$r_{\rm{drag}}$ [Mpc] & $147.17_{-0.25-0.51}^{+0.26+0.51}$ & $147.53_{-0.21-0.40}^{+0.21+0.42}$ & $147.58_{-0.19-0.39}^{+0.20+0.38}$ & $147.61_{-0.20-0.40}^{+0.20+0.40}$ & $147.60_{-0.20-0.38}^{+0.19+0.40}$ & $147.59_{-0.20-0.38}^{+0.20+0.38}$ \\
				
				\hline
				
				$\Delta\chi^2_{\rm min}$ & $-0.9$ & $0.9$ & $-2.0$ & $-3.9$ & $-3.8$ & $-2.1$  \\
				
				$\ln \mathcal{B}_{ij}$ & $1.5$ & $-1.4$ & $-0.7$ & $0.6$ & $0.0$ & $-0.4$ \\

		\end{tabular}}
        \scalebox{0.85}{
        \begin{tabular}{cccccccc}
			\hline
			Parameters & CMB & CMB+DESI & CMB+DESI+PantheonPlus & CMB+DESI+Union3 & CMB+DESI+DES-Dovekie & CMB+DESI+Union3.1  \\ \hline

            $\Omega_\mathrm{b} h^2$ & $0.02248_{-0.00018-0.00034}^{+0.00017+0.00036}$ & $0.02246_{-0.00013-0.00027}^{+0.00014+0.00026}$ & $0.02245_{-0.00013-0.00026}^{+0.00013+0.00026}$ & $0.02247_{-0.00013-0.00026}^{+0.00013+0.00026}$ & $0.02246_{-0.00013-0.00026}^{+0.00013+0.00026}$ & $0.02245_{-0.00014-0.00026}^{+0.00014+0.00027}$ \\
				
				$\Omega_\mathrm{c} h^2$ & $0.1188_{-0.0016-0.0030}^{+0.0015+0.0031}$ & $0.11890_{-0.00089-0.0017}^{+0.00090+0.0018}$ & $0.11904_{-0.00089-0.0017}^{+0.00089+0.0017}$ & $0.11887_{-0.00087-0.0017}^{+0.00087+0.0017}$ & $0.11899_{-0.00088-0.0016}^{+0.00086+0.0017}$ & $0.11897_{-0.00088-0.0018}^{+0.00086+0.0017}$ \\
				
				$100\theta_\mathrm{MC}$ & $1.04191_{-0.00029-0.00060}^{+0.00030+0.00059}$ & $1.04191_{-0.00028-0.00055}^{+0.00028+0.00055}$ & $1.04188_{-0.00028-0.00057}^{+0.00029+0.00055}$ & $1.04192_{-0.00027-0.00054}^{+0.00028+0.00055}$ & $1.04189_{-0.00028-0.00056}^{+0.00028+0.00055}$ & $1.04191_{-0.00026-0.00059}^{+0.00030+0.00052}$ \\
				
				$\tau$ & $0.0509_{-0.0078-0.016}^{+0.0078+0.016}$ & $0.0520_{-0.0080-0.016}^{+0.0080+0.016}$ & $0.0518_{-0.0078-0.016}^{+0.0077+0.015}$ & $0.0514_{-0.0071-0.015}^{+0.0071+0.016}$ & $0.0514_{-0.0075-0.016}^{+0.0076+0.015}$ & $0.0513_{-0.0074-0.017}^{+0.0079+0.015}$ \\
				
				$n_\mathrm{s}$ & $0.9677_{-0.0051-0.0096}^{+0.0047+0.010}$ & $0.9676_{-0.0037-0.0073}^{+0.0036+0.0073}$ & $0.9670_{-0.0036-0.0070}^{+0.0036+0.0070}$ & $0.9675_{-0.0035-0.0068}^{+0.0034+0.0068}$ & $0.9669_{-0.0034-0.0073}^{+0.0037+0.0070}$ & $0.9673_{-0.0036-0.0070}^{+0.0035+0.0070}$ \\
				
				$\ln(10^{10} A_\mathrm{s})$ & $3.034_{-0.016-0.032}^{+0.016+0.032}$ & $3.037_{-0.016-0.033}^{+0.016+0.031}$ & $3.037_{-0.015-0.031}^{+0.015+0.029}$ & $3.036_{-0.015-0.028}^{+0.014+0.031}$ & $3.036_{-0.015-0.031}^{+0.015+0.029}$ & $3.036_{-0.015-0.032}^{+0.015+0.031}$ \\
				
			\rowcolor{LightSkyBlue}	$\sum m_{\nu,{\rm eff}}$ & $-0.13_{-0.16-0.34}^{+0.16+0.33}$ & $-0.131_{-0.064-0.13}^{+0.063+0.14}$ & $-0.119_{-0.064-0.11}^{+0.054+0.13}$ & $-0.141_{-0.058-0.11}^{+0.057+0.12}$ & $-0.126_{-0.055-0.11}^{+0.054+0.12}$ & $-0.131_{-0.059-0.12}^{+0.057+0.13}$ \\
				
				$w_0$ & $<-1.49<-0.15$ & $-0.77_{-0.19-0.38}^{+0.19+0.36}$ & $-0.866_{-0.056-0.11}^{+0.056+0.11}$ & $-0.718_{-0.086-0.17}^{+0.088+0.18}$ & $-0.836_{-0.053-0.11}^{+0.054+0.11}$ & $-0.797_{-0.084-0.16}^{+0.076+0.16}$ \\
				
	$\Omega_\mathrm{m}$ & $0.222_{-0.095-0.11}^{+0.032+0.17}$ & $0.314_{-0.017-0.033}^{+0.017+0.033}$ & $0.3053_{-0.0057-0.011}^{+0.0058+0.012}$ & $0.3177_{-0.0084-0.016}^{+0.0082+0.017}$ & $0.3074_{-0.0055-0.011}^{+0.0056+0.011}$ & $0.3109_{-0.0077-0.015}^{+0.0078+0.015}$ \\
				
	$\sigma_8$ & $0.944_{-0.091-0.20}^{+0.13+0.18}$ & $0.820_{-0.019-0.035}^{+0.019+0.038}$ & $0.828_{-0.013-0.026}^{+0.013+0.025}$ & $0.816_{-0.014-0.027}^{+0.014+0.028}$ & $0.825_{-0.013-0.025}^{+0.013+0.025}$ & $0.823_{-0.013-0.027}^{+0.014+0.028}$ \\
				
	$H_0$ [km/s/Mpc] & $82_{-11-25}^{+16+21}$ & $66.8_{-2.0-3.6}^{+1.7+3.8}$ & $67.70_{-0.60-1.2}^{+0.59+1.2}$ & $66.26_{-0.88-1.7}^{+0.88+1.7}$ & $67.44_{-0.58-1.1}^{+0.56+1.1}$ & $67.03_{-0.83-1.6}^{+0.81+1.6}$ \\
				
	$S_8$ & $0.785_{-0.054-0.072}^{+0.035+0.089}$ & $0.838_{-0.014-0.026}^{+0.013+0.026}$ & $0.835_{-0.012-0.022}^{+0.012+0.023}$ & $0.840_{-0.012-0.023}^{+0.012+0.023}$ & $0.835_{-0.011-0.024}^{+0.012+0.023}$ & $0.837_{-0.011-0.023}^{+0.012+0.022}$ \\
				
	$r_{\rm{drag}}$ [Mpc] & $147.33_{-0.32-0.65}^{+0.33+0.64}$ & $147.31_{-0.22-0.43}^{+0.22+0.45}$ & $147.29_{-0.22-0.44}^{+0.22+0.42}$ & $147.31_{-0.22-0.46}^{+0.22+0.43}$ & $147.29_{-0.20-0.44}^{+0.23+0.40}$ & $147.30_{-0.22-0.44}^{+0.21+0.44}$ \\
				
				\hline
				
	$\Delta\chi^2_{\rm min}$ & $-0.6$ & $-0.7$ & $-4.9$ & $-8.7$ & $-7.8$ & $-5.3$  \\
				
	$\ln \mathcal{B}_{ij}$ & $1.4$ & $-0.4$ & $0.9$ & $3.0$ & $2.0$ & $1.3$  \\

			\hline                                                        
		\end{tabular}}                                                       
		\caption{As Table~\ref{table:generalized-thawing-p-1}, but for $p=5$ {\bf generalized thawing dark energy}.}
        
		\label{table:generalized-thawing-p-5}                   
	\end{table*}                                     
\end{center}
\endgroup

\begingroup
\squeezetable                                   
\begin{center}  
	\begin{table*}
    \scalebox{0.85}{
		\begin{tabular}{cccccccc}
			\hline
			Parameters & CMB & CMB+DESI & CMB+DESI+PantheonPlus & CMB+DESI+Union3 & CMB+DESI+DES-Dovekie  & CMB+DESI+Union3.1 \\ \hline

           $\Omega_\mathrm{b} h^2$ & $0.02236_{-0.00015-0.00029}^{+0.00015+0.00029}$ & $0.02254_{-0.00013-0.00026}^{+0.00013+0.00026}$ & $0.02254_{-0.00013-0.00025}^{+0.00012+0.00025}$ & $0.02255_{-0.00013-0.00025}^{+0.00013+0.00025}$ & $0.02255_{-0.00012-0.00025}^{+0.00012+0.00025}$ & $0.02255_{-0.00012-0.00025}^{+0.00013+0.00026}$ \\
				
				$\Omega_\mathrm{c} h^2$ & $0.1199_{-0.0012-0.0024}^{+0.0012+0.0024}$ & $0.11754_{-0.00073-0.0014}^{+0.00073+0.0015}$ & $0.11756_{-0.00064-0.0013}^{+0.00065+0.0013}$ & $0.11741_{-0.00067-0.0013}^{+0.00067+0.0013}$ & $0.11748_{-0.00067-0.0013}^{+0.00066+0.0014}$ & $0.11741_{-0.00067-0.0014}^{+0.00068+0.0013}$ \\

        $100\theta_\mathrm{MC}$ & $1.04186_{-0.00029-0.00057}^{+0.00029+0.00058}$ & $1.04208_{-0.00029-0.00054}^{+0.00028+0.00056}$ & $1.04209_{-0.00029-0.00054}^{+0.00027+0.00055}$ & $1.04213_{-0.00028-0.00055}^{+0.00028+0.00056}$ & $1.04210_{-0.00027-0.00055}^{+0.00028+0.00053}$ & $1.04210_{-0.00026-0.00053}^{+0.00026+0.00053}$ \\

        $\tau$ & $0.0539_{-0.0072-0.015}^{+0.0071+0.015}$ & $0.0610_{-0.0078-0.015}^{+0.0071+0.015}$ & $0.0607_{-0.0079-0.014}^{+0.0068+0.015}$ & $0.0614_{-0.0080-0.014}^{+0.0070+0.016}$ & $0.0611_{-0.0080-0.014}^{+0.0071+0.015}$ & $0.0614_{-0.0078-0.015}^{+0.0072+0.015}$ \\
				
				$n_\mathrm{s}$ & $0.9648_{-0.0042-0.0083}^{+0.0042+0.0080}$ & $0.9707_{-0.0036-0.0070}^{+0.0035+0.0069}$ & $0.9706_{-0.0033-0.0064}^{+0.0033+0.0064}$ & $0.9710_{-0.0034-0.0067}^{+0.0034+0.0066}$ & $0.9707_{-0.0034-0.0067}^{+0.0034+0.0067}$ & $0.9709_{-0.0033-0.0065}^{+0.0034+0.0068}$ \\
				
		$\ln(10^{10} A_\mathrm{s})$ & $3.043_{-0.014-0.029}^{+0.014+0.029}$ & $3.053_{-0.015-0.029}^{+0.015+0.030}$ & $3.053_{-0.016-0.028}^{+0.014+0.031}$ & $3.054_{-0.016-0.028}^{+0.014+0.031}$ & $3.053_{-0.016-0.028}^{+0.014+0.030}$ & $3.053_{-0.016-0.028}^{+0.014+0.031}$ \\
				
		$w_0$ & $<-0.94<0.17$ & $-0.86_{-0.56-1.4}^{+0.78+1.3}$ & $-0.81_{-0.11-0.21}^{+0.10+0.21}$ & $-0.47_{-0.20-0.39}^{+0.20+0.38}$ & $-0.733_{-0.098-0.21}^{+0.10+0.20}$ & $-0.61_{-0.18-0.35}^{+0.18+0.37}$ \\
				
	$\Omega_\mathrm{m}$ &
$0.292_{-0.072-0.077}^{+0.034+0.097}$ & 
$0.309_{-0.033-0.064}^{+0.035+0.066}$ & 
$0.3103_{-0.0060-0.011}^{+0.0060+0.012}$ & 
$0.328_{-0.011-0.021}^{+0.010+0.021}$ & 
$0.3140_{-0.0057-0.012}^{+0.0055+0.011}$ & 
$0.3200_{-0.0095-0.018}^{+0.0097+0.019}$ \\

$\sigma_8$ & 
$0.837_{-0.041-0.090}^{+0.072+0.082}$ & 
$0.801_{-0.038-0.061}^{+0.029+0.066}$ & 
$0.798_{-0.008-0.016}^{+0.008+0.015}$ & 
$0.781_{-0.011-0.021}^{+0.011+0.021}$ & 
$0.794_{-0.008-0.016}^{+0.008+0.016}$ & 
$0.788_{-0.011-0.021}^{+0.010+0.022}$ \\

$H_0$ [km/s/Mpc] & 
$70.7_{-7.5-10}^{+8.8+9.9}$ & 
$67.8_{-4.8-7.0}^{+2.7+7.3}$ & 
$67.36_{-0.60-1.2}^{+0.61+1.2}$ & 
$65.5_{-1.1-2.0}^{+1.0+2.1}$ & 
$66.94_{-0.59-1.1}^{+0.54+1.2}$ & 
$66.30_{-1.0-1.9}^{+0.98+2.0}$ \\
				
				$S_8$ & $0.819_{-0.026-0.045}^{+0.027+0.046}$ & $0.810_{-0.012-0.028}^{+0.014+0.025}$ & $0.811_{-0.008-0.016}^{+0.008+0.017}$ & $0.817_{-0.009-0.017}^{+0.009+0.017}$ & $0.812_{-0.008-0.016}^{+0.008+0.016}$ & $0.814_{-0.009-0.017}^{+0.009+0.017}$ \\
				
				$r_{\rm{drag}}$ [Mpc] & $147.13_{-0.26-0.52}^{+0.26+0.53}$ & $147.57_{-0.21-0.42}^{+0.21+0.40}$ & $147.55_{-0.19-0.37}^{+0.19+0.39}$ & $147.59_{-0.20-0.38}^{+0.19+0.37}$ & $147.57_{-0.19-0.39}^{+0.20+0.39}$ & $147.59_{-0.19-0.38}^{+0.20+0.39}$ \\
				
				\hline
				
				$\Delta\chi^2_{\rm min}$ & $0.1$ & $0.7$ & $-2.4$ & $-6.2$ & $-5.6$ & $-3.7$  \\
				
				$\ln \mathcal{B}_{ij}$ & $0.9$ & $0.1$ & $0.0$ & $3.0$ & $1.6$ & $1.2$  \\

		\end{tabular}}
        \scalebox{0.85}{
        \begin{tabular}{cccccccc}
			\hline
			Parameters & CMB & CMB+DESI & CMB+DESI+PantheonPlus & CMB+DESI+Union3 & CMB+DESI+DES-Dovekie & CMB+DESI+Union3.1  \\ \hline

				$\Omega_\mathrm{b} h^2$ & $0.02248_{-0.00017-0.00035}^{+0.00018+0.00035}$ & $0.02246_{-0.00013-0.00027}^{+0.00013+0.00025}$ & $0.02244_{-0.00013-0.00026}^{+0.00013+0.00026}$ & $0.02245_{-0.00013-0.00025}^{+0.00013+0.00026}$ & $0.02244_{-0.00013-0.00026}^{+0.00013+0.00026}$ & $0.02245_{-0.00013-0.00026}^{+0.00013+0.00025}$ \\
				
				$\Omega_\mathrm{c} h^2$ & $0.1188_{-0.0015-0.0029}^{+0.0015+0.0030}$ & $0.11892_{-0.00091-0.0017}^{+0.00086+0.0017}$ & $0.11909_{-0.00086-0.0017}^{+0.00086+0.0017}$ & $0.11898_{-0.00091-0.0018}^{+0.00089+0.0017}$ & $0.11911_{-0.00093-0.0017}^{+0.00092+0.0018}$ & $0.11903_{-0.00089-0.0017}^{+0.00089+0.0017}$ \\

$100\theta_\mathrm{MC}$ & $1.04194_{-0.00030-0.00057}^{+0.00031+0.00058}$ & $1.04190_{-0.00030-0.00057}^{+0.00029+0.00057}$ & $1.04190_{-0.00028-0.00056}^{+0.00028+0.00055}$ & $1.04192_{-0.00028-0.00057}^{+0.00029+0.00055}$ & $1.04190_{-0.00027-0.00056}^{+0.00029+0.00054}$ & $1.04190_{-0.00028-0.00055}^{+0.00028+0.00057}$ \\

$\tau$ & $0.0514_{-0.0077-0.016}^{+0.0080+0.015}$ & $0.0521_{-0.0076-0.015}^{+0.0075+0.015}$ & $0.0514_{-0.0074-0.016}^{+0.0076+0.015}$ & $0.0519_{-0.0077-0.016}^{+0.0078+0.016}$ & $0.0517_{-0.0074-0.016}^{+0.0075+0.015}$ & $0.0517_{-0.0074-0.015}^{+0.0073+0.015}$ \\
				
				$n_\mathrm{s}$ & $0.9679_{-0.0048-0.0095}^{+0.0049+0.0099}$ & $0.9675_{-0.0035-0.0073}^{+0.0035+0.0069}$ & $0.9670_{-0.0035-0.0069}^{+0.0035+0.0068}$ & $0.9675_{-0.0034-0.0071}^{+0.0036+0.0071}$ & $0.9669_{-0.0037-0.0070}^{+0.0036+0.0069}$ & $0.9670_{-0.0035-0.0071}^{+0.0035+0.0072}$ \\
				
				$\ln(10^{10} A_\mathrm{s})$ & $3.035_{-0.016-0.034}^{+0.016+0.032}$ & $3.037_{-0.015-0.030}^{+0.015+0.029}$ & $3.036_{-0.014-0.030}^{+0.015+0.029}$ & $3.037_{-0.015-0.031}^{+0.015+0.031}$ & $3.037_{-0.015-0.031}^{+0.015+0.030}$ & $3.037_{-0.015-0.030}^{+0.015+0.030}$ \\
				
			\rowcolor{LightSkyBlue}	$\sum m_{\nu,{\rm eff}}$ & $-0.14_{-0.17-0.34}^{+0.17+0.33}$ & $-0.114_{-0.059-0.12}^{+0.058+0.13}$ & $-0.104_{-0.065-0.11}^{+0.053+0.13}$ & $-0.114_{-0.064-0.11}^{+0.052+0.13}$ & $-0.113_{-0.063-0.12}^{+0.056+0.12}$ & $-0.115_{-0.063-0.11}^{+0.051+0.13}$ \\
				
				$w_0$ & $<-0.98<0.15$ & $-0.33_{-0.54-1.1}^{+0.49+0.91}$ & $-0.77_{-0.10-0.20}^{+0.10+0.19}$ & $-0.36_{-0.20-0.39}^{+0.20+0.38}$ & $-0.67_{-0.11-0.20}^{+0.11+0.21}$ & $-0.53_{-0.18-0.35}^{+0.18+0.37}$ \\

$\Omega_\mathrm{m}$ & $0.266_{-0.068-0.091}^{+0.041+0.11}$ & $0.327_{-0.028-0.052}^{+0.041+0.050}$ & $0.3052_{-0.0059-0.012}^{+0.0062+0.012}$ & $0.325_{-0.010-0.020}^{+0.010+0.020}$ & $0.3096_{-0.0060-0.012}^{+0.0061+0.012}$ & $0.3168_{-0.0098-0.017}^{+0.0086+0.020}$ \\

$\sigma_8$ & $0.875_{-0.069-0.12}^{+0.065+0.12}$ & $0.807_{-0.032-0.052}^{+0.023+0.057}$ & $0.827_{-0.014-0.027}^{+0.013+0.026}$ & $0.809_{-0.016-0.030}^{+0.016+0.029}$ & $0.824_{-0.014-0.026}^{+0.014+0.027}$ & $0.817_{-0.015-0.030}^{+0.015+0.028}$ \\

$H_0$ [km/s/Mpc] & $73.3_{-7.8-12}^{+7.3+12}$ & $65.5_{-4.3-4.8}^{+2.1+5.7}$ & $67.77_{-0.63-1.2}^{+0.59+1.2}$ & $65.6_{-1.1-2.0}^{+1.0+2.1}$ & $67.27_{-0.60-1.2}^{+0.61+1.2}$ & $66.49_{-0.96-2.0}^{+0.96+1.9}$ \\
				
	$S_8$ & $0.814_{-0.027-0.050}^{+0.029+0.049}$ & $0.842_{-0.014-0.031}^{+0.017+0.029}$ & $0.834_{-0.012-0.022}^{+0.012+0.023}$ & $0.842_{-0.012-0.024}^{+0.012+0.023}$ & $0.837_{-0.013-0.023}^{+0.012+0.023}$ & $0.839_{-0.012-0.024}^{+0.012+0.023}$ \\
				
$r_{\rm{drag}}$ [Mpc] & $147.34_{-0.33-0.64}^{+0.32+0.64}$ & $147.30_{-0.22-0.43}^{+0.22+0.44}$ & $147.27_{-0.21-0.43}^{+0.22+0.42}$ & $147.29_{-0.22-0.43}^{+0.23+0.45}$ & $147.27_{-0.22-0.44}^{+0.23+0.43}$ & $147.28_{-0.22-0.44}^{+0.22+0.45}$ \\
				
				\hline
				
	$\Delta\chi^2_{\rm min}$ & $-0.2$ & $-1.4$ & $-4.2$ & $-9.1$ & $-8.4$ & $-6.0$  \\
				
	$\ln \mathcal{B}_{ij}$ & $1.4$ & $0.9$ & $1.1$ & $4.4$ & $3.0$ & $2.5$  \\

			\hline                                                        
		\end{tabular}}                                                       
		\caption{As Table~\ref{table:generalized-thawing-p-1}, but for $p=15$ {\bf generalized thawing dark energy}.}
        
		\label{table:generalized-thawing-p-15}                   
	\end{table*}                                     
\end{center}
\endgroup

\begingroup
\squeezetable                                   
\begin{center}  
	\begin{table*}
    \scalebox{0.85}{
		\begin{tabular}{cccccccc}
			\hline
			Parameters & CMB & CMB+DESI & CMB+DESI+PantheonPlus & CMB+DESI+Union3 & CMB+DESI+DES-Dovekie & CMB+DESI+Union3.1 \\ \hline

            $\Omega_\mathrm{b} h^2$ & $0.02242_{-0.00016-0.00030}^{+0.00015+0.00031}$ & $0.02238_{-0.00014-0.00028}^{+0.00014+0.00027}$ & $0.02243_{-0.00013-0.00026}^{+0.00013+0.00026}$ & $0.02236_{-0.00013-0.00026}^{+0.00013+0.00025}$ & $0.02241_{-0.00013-0.00025}^{+0.00013+0.00025}$ & $0.02239_{-0.00013-0.00026}^{+0.00013+0.00025}$\\
			
			$\Omega_\mathrm{c} h^2$ & $0.1192_{-0.0013-0.0024}^{+0.0013+0.0025}$ & $0.11963_{-0.00095-0.0019}^{+0.00097+0.0019}$ & $0.11896_{-0.00071-0.0014}^{+0.00072+0.0014}$ & $0.11981_{-0.00080-0.0015}^{+0.00078+0.0015}$ & $0.11916_{-0.00066-0.0014}^{+0.00071+0.0013}$ & $0.11950_{-0.00077-0.0015}^{+0.00077+0.0015}$\\

$100\theta_\mathrm{MC}$ & $1.04192_{-0.00030-0.00059}^{+0.00031+0.00058}$ & $1.04189_{-0.00029-0.00056}^{+0.00028+0.00056}$ & $1.04198_{-0.00027-0.00053}^{+0.00028+0.00054}$ & $1.04187_{-0.00028-0.00057}^{+0.00028+0.00055}$ & $1.04193_{-0.00028-0.00055}^{+0.00028+0.00054}$ & $1.04191_{-0.00027-0.00054}^{+0.00028+0.00055}$\\

$\tau$ & $0.0519_{-0.0072-0.014}^{+0.0072+0.015}$ & $0.0531_{-0.0073-0.014}^{+0.0073+0.015}$ & $0.0555_{-0.0071-0.014}^{+0.0070+0.015}$ & $0.0522_{-0.0068-0.014}^{+0.0069+0.014}$ & $0.0545_{-0.0068-0.014}^{+0.0067+0.014}$ & $0.0534_{-0.0069-0.013}^{+0.0067+0.015}$\\
			
			$n_\mathrm{s}$ & $0.9665_{-0.0046-0.0083}^{+0.0042+0.0088}$ & $0.9654_{-0.0037-0.0072}^{+0.0038+0.0071}$ & $0.9670_{-0.0033-0.0065}^{+0.0033+0.0066}$ & $0.9650_{-0.0034-0.0067}^{+0.0034+0.0067}$ & $0.9666_{-0.0033-0.0068}^{+0.0033+0.0066}$ & $0.9657_{-0.0035-0.0066}^{+0.0035+0.0070}$\\
			
			$\ln(10^{10} A_\mathrm{s})$ & $3.037_{-0.014-0.028}^{+0.014+0.028}$ & $3.041_{-0.014-0.029}^{+0.014+0.029}$ & $3.044_{-0.014-0.028}^{+0.014+0.029}$ & $3.039_{-0.014-0.027}^{+0.014+0.028}$ & $3.043_{-0.014-0.028}^{+0.014+0.028}$ & $3.041_{-0.014-0.027}^{+0.014+0.029}$\\
		
			$w_0$ & $-0.18_{-0.30-0.79}^{+0.73+0.75}$ & $-0.74_{-0.13-0.22}^{+0.10+0.24}$ & $-0.845_{-0.056-0.098}^{+0.051+0.11}$ & $-0.721_{-0.089-0.16}^{+0.080+0.17}$ & $-0.813_{-0.056-0.11}^{+0.055+0.11}$ & $-0.763_{-0.076-0.15}^{+0.075+0.15}$\\

$\Omega_\mathrm{m}$ & $0.339_{-0.025-0.033}^{+0.015+0.039}$ & $0.3176_{-0.0099-0.016}^{+0.0073+0.018}$ & $0.3105_{-0.0050-0.0092}^{+0.0046+0.0096}$ & $0.3191_{-0.0067-0.013}^{+0.0066+0.014}$ & $0.3125_{-0.0048-0.0095}^{+0.0048+0.0097}$ & $0.3159_{-0.0065-0.011}^{+0.0058+0.013}$\\

$\sigma_8$ & $0.789_{-0.026-0.034}^{+0.023+0.027}$ & $0.809_{-0.006-0.012}^{+0.006+0.012}$ & $0.809_{-0.006-0.011}^{+0.006+0.012}$ & $0.809_{-0.006-0.011}^{+0.006+0.012}$ & $0.809_{-0.006-0.012}^{+0.006+0.012}$ & $0.809_{-0.006-0.011}^{+0.006+0.011}$\\

$H_0$ [km/s/Mpc] & $64.8_{-3.0-3.7}^{+2.7+3.1}$ & $67.03_{-0.60-1.5}^{+0.83+1.4}$ & $67.64_{-0.38-0.77}^{+0.38+0.74}$ & $66.90_{-0.54-1.1}^{+0.56+1.0}$ & $67.47_{-0.39-0.78}^{+0.38+0.77}$ & $67.18_{-0.47-1.0}^{+0.53+0.95}$\\
		
			$S_8$ & $0.838_{-0.014-0.027}^{+0.014+0.027}$ & $0.832_{-0.012-0.023}^{+0.012+0.025}$ & $0.823_{-0.009-0.018}^{+0.009+0.018}$ & $0.834_{-0.010-0.020}^{+0.010+0.020}$ & $0.826_{-0.009-0.017}^{+0.009+0.018}$ & $0.830_{-0.010-0.019}^{+0.010+0.020}$\\
		
			$r_{\rm{drag}}$ [Mpc] & $147.25_{-0.28-0.54}^{+0.28+0.54}$ & $147.18_{-0.23-0.44}^{+0.23+0.45}$ & $147.30_{-0.20-0.40}^{+0.20+0.39}$ & $147.16_{-0.21-0.40}^{+0.21+0.40}$ & $147.27_{-0.21-0.38}^{+0.20+0.40}$ & $147.21_{-0.21-0.40}^{+0.20+0.41}$\\
			
			\hline
			
			$\Delta\chi^2_{\rm min}$ & $-1.8$ & $-5.7$ & $-8.2$ & $-12.6$ & $-11.3$ & $-10.3$  \\
			
			$\ln \mathcal{B}_{ij}$ & $1.3$ & $1.4$ & $1.8$ & $4.9$ & $3.8$ & $3.7$  \\
			
            \end{tabular}}
             \scalebox{0.85}{
			\begin{tabular}{cccccccc}
			\hline
			Parameters & CMB & CMB+DESI & CMB+DESI+PantheonPlus & CMB+DESI+Union3 & CMB+DESI+DES-Dovekie  & CMB+DESI+Union3.1 \\ \hline
			
			$\Omega_\mathrm{b} h^2$ & $0.02248_{-0.00018-0.00035}^{+0.00017+0.00036}$ & $0.02238_{-0.00014-0.00027}^{+0.00014+0.00029}$ & $0.02240_{-0.00013-0.00026}^{+0.00013+0.00025}$ & $0.02237_{-0.00013-0.00026}^{+0.00013+0.00024}$ & $0.02240_{-0.00013-0.00026}^{+0.00013+0.00026}$ & $0.02238_{-0.00013-0.00026}^{+0.00013+0.00026}$\\
			
			$\Omega_\mathrm{c} h^2$ & $0.1186_{-0.0016-0.0031}^{+0.0016+0.0031}$ & $0.11964_{-0.00095-0.0019}^{+0.00095+0.0019}$ & $0.11949_{-0.00086-0.0017}^{+0.00084+0.0017}$ & $0.11987_{-0.00087-0.0017}^{+0.00087+0.0017}$ & $0.11963_{-0.00083-0.0017}^{+0.00084+0.0017}$ & $0.11971_{-0.00088-0.0018}^{+0.00088+0.0017}$\\
			
$100\theta_\mathrm{MC}$ & $1.04193_{-0.00030-0.00059}^{+0.00029+0.00060}$ & $1.04188_{-0.00028-0.00055}^{+0.00029+0.00055}$ & $1.04187_{-0.00029-0.00055}^{+0.00028+0.00057}$ & $1.04186_{-0.00026-0.00056}^{+0.00028+0.00053}$ & $1.04188_{-0.00029-0.00057}^{+0.00029+0.00057}$ & $1.04187_{-0.00028-0.00056}^{+0.00028+0.00055}$\\
			
$\tau$ & $0.0507_{-0.0078-0.016}^{+0.0081+0.017}$ & $0.0519_{-0.0075-0.015}^{+0.0075+0.016}$ & $0.0520_{-0.0074-0.015}^{+0.0074+0.016}$ & $0.0516_{-0.0073-0.015}^{+0.0073+0.015}$ & $0.0523_{-0.0076-0.015}^{+0.0072+0.015}$ & $0.0519_{-0.0072-0.015}^{+0.0072+0.015}$\\
			
			$n_\mathrm{s}$ & $0.9681_{-0.0051-0.010}^{+0.0051+0.0099}$ & $0.9655_{-0.0038-0.0071}^{+0.0035+0.0074}$ & $0.9658_{-0.0035-0.0066}^{+0.0035+0.0070}$ & $0.9649_{-0.0035-0.0067}^{+0.0035+0.0069}$ & $0.9656_{-0.0036-0.0070}^{+0.0036+0.0068}$ & $0.9653_{-0.0035-0.0070}^{+0.0035+0.0069}$\\
			
			$\ln(10^{10} A_\mathrm{s})$ & $3.033_{-0.016-0.034}^{+0.017+0.033}$ & $3.038_{-0.015-0.030}^{+0.015+0.031}$ & $3.038_{-0.015-0.029}^{+0.015+0.031}$ & $3.038_{-0.014-0.029}^{+0.014+0.029}$ & $3.039_{-0.014-0.030}^{+0.015+0.029}$ & $3.039_{-0.015-0.029}^{+0.014+0.029}$\\

\rowcolor{LightSkyBlue}         
        $\sum m_{\nu,\rm{eff}}$ & $-0.05_{-0.16-0.33}^{+0.16+0.35}$ & $0.00_{-0.10-0.22}^{+0.12+0.21}$ & $-0.010_{-0.068-0.13}^{+0.069+0.13}$ & $0.055_{-0.066-0.14}^{+0.073+0.13}$ & $0.006_{-0.064-0.13}^{+0.069+0.12}$ & $0.024_{-0.069-0.14}^{+0.076+0.13}$\\
			
$w_0$ & $-0.29_{-0.38-0.83}^{+0.82+0.85}$ & $-0.81_{-0.21-0.34}^{+0.14+0.39}$ & $-0.871_{-0.061-0.11}^{+0.054+0.12}$ & $-0.726_{-0.10-0.19}^{+0.099+0.21}$ & $-0.833_{-0.061-0.12}^{+0.060+0.12}$ & $-0.791_{-0.089-0.17}^{+0.089+0.18}$\\
			
$\Omega_\mathrm{m}$ & $0.319_{-0.042-0.064}^{+0.030+0.078}$ & $0.311_{-0.018-0.031}^{+0.014+0.036}$ & $0.3063_{-0.0060-0.011}^{+0.0059+0.012}$ & $0.3186_{-0.0090-0.018}^{+0.0090+0.019}$ & $0.3093_{-0.0060-0.012}^{+0.0060+0.012}$ & $0.3127_{-0.0080-0.016}^{+0.0081+0.017}$\\
			
$\sigma_8$ & $0.812_{-0.036-0.082}^{+0.042+0.076}$ & $0.820_{-0.021-0.047}^{+0.023+0.043}$ & $0.823_{-0.013-0.026}^{+0.013+0.027}$ & $0.810_{-0.014-0.030}^{+0.016+0.028}$ & $0.821_{-0.013-0.026}^{+0.013+0.028}$ & $0.817_{-0.014-0.030}^{+0.014+0.029}$\\
			
$H_0$ [km/s/Mpc] & $66.4_{-2.8-6.3}^{+3.7+5.7}$ & $67.6_{-1.2-2.9}^{+1.6+2.7}$ & $68.04_{-0.51-1.0}^{+0.51+0.98}$ & $66.97_{-0.77-1.6}^{+0.78+1.5}$ & $67.78_{-0.51-1.0}^{+0.51+1.0}$ & $67.47_{-0.69-1.5}^{+0.70+1.4}$\\
			
			$S_8$ & $0.834_{-0.015-0.029}^{+0.015+0.029}$ & $0.833_{-0.012-0.024}^{+0.012+0.024}$ & $0.831_{-0.011-0.022}^{+0.011+0.023}$ & $0.834_{-0.012-0.023}^{+0.012+0.023}$ & $0.833_{-0.011-0.023}^{+0.011+0.023}$ & $0.833_{-0.012-0.025}^{+0.012+0.023}$\\
			
			$r_{\rm{drag}}$ [Mpc] & $147.35_{-0.33-0.63}^{+0.33+0.65}$ & $147.18_{-0.24-0.45}^{+0.22+0.47}$ & $147.21_{-0.21-0.43}^{+0.21+0.43}$ & $147.13_{-0.22-0.42}^{+0.22+0.44}$ & $147.17_{-0.23-0.44}^{+0.22+0.45}$ & $147.17_{-0.22-0.42}^{+0.22+0.42}$\\
			
			\hline
			
			$\Delta\chi^2_{\rm min}$ & $-0.7$ & $0.4$ & $-4.1$ & $-7.5$ & $-7.0$ & $-4.7$ \\
			
			$\ln \mathcal{B}_{ij}$ & $0.6$ & $-0.7$ & $0.2$ & $2.6$ & $1.8$ & $1.1$ \\

			\hline                                                       
		\end{tabular}}                                                       
		\caption{Summary of the observational constraints on various free and derived cosmological parameters of 
{\bf mirage dark energy} at 68\% and 95\% CL using several observational datasets. 
The upper half of the table corresponds to the scenario with neutrino mass fixed, while the lower half of the table corresponds to the scenario including neutrino mass as a free parameter.}
		\label{table:mirage}                   
	\end{table*}                                     
\end{center}
\endgroup

\begingroup
\squeezetable                                   
\begin{center}  
	\begin{table*}
    \scalebox{0.85}{
		\begin{tabular}{cccccccc}
			\hline
			Parameters & CMB & CMB+DESI & CMB+DESI+PantheonPlus & CMB+DESI+Union3 & CMB+DESI+DES-Dovekie & CMB+DESI+Union3.1 \\ \hline
            	
		$\Omega_\mathrm{b} h^2$ & $0.02238_{-0.00014-0.00030}^{+0.00015+0.00030}$ & $0.02247_{-0.00013-0.00026}^{+0.00013+0.00025}$ & $0.02249_{-0.00013-0.00025}^{+0.00013+0.00024}$ & $0.02248_{-0.00013-0.00026}^{+0.00013+0.00025}$ & $0.02248_{-0.00012-0.00025}^{+0.00012+0.00024}$ & $0.02249_{-0.00013-0.00026}^{+0.00013+0.00025}$ \\

$\Omega_\mathrm{c} h^2$ & $0.1196_{-0.0012-0.0023}^{+0.0012+0.0024}$ & $0.11847_{-0.00086-0.0015}^{+0.00077+0.0017}$ & $0.11820_{-0.00071-0.0013}^{+0.00066+0.0014}$ & $0.11821_{-0.00072-0.0013}^{+0.00066+0.0014}$ & $0.11821_{-0.00068-0.0013}^{+0.00067+0.0013}$ & $0.11819_{-0.00071-0.0014}^{+0.00069+0.0014}$ \\

$100\theta_\mathrm{MC}$ & $1.04191_{-0.00030-0.00058}^{+0.00030+0.00061}$ & $1.04201_{-0.00028-0.00053}^{+0.00028+0.00055}$ & $1.04204_{-0.00028-0.00056}^{+0.00028+0.00054}$ & $1.04203_{-0.00028-0.00056}^{+0.00028+0.00054}$ & $1.04206_{-0.00028-0.00055}^{+0.00028+0.00055}$ & $1.04206_{-0.00028-0.00055}^{+0.00028+0.00054}$ \\

$\tau$ & $0.0536_{-0.0074-0.014}^{+0.0069+0.015}$ & $0.0575_{-0.0079-0.014}^{+0.0068+0.016}$ & $0.0587_{-0.0075-0.014}^{+0.0068+0.015}$ & $0.0588_{-0.0077-0.014}^{+0.0068+0.014}$ & $0.0585_{-0.0074-0.014}^{+0.0068+0.015}$ & $0.0586_{-0.0079-0.014}^{+0.0067+0.015}$ \\

$n_\mathrm{s}$ & $0.9654_{-0.0041-0.0081}^{+0.0041+0.0083}$ & $0.9683_{-0.0035-0.0070}^{+0.0035+0.0070}$ & $0.9691_{-0.0033-0.0066}^{+0.0033+0.0064}$ & $0.9691_{-0.0034-0.0066}^{+0.0034+0.0066}$ & $0.9689_{-0.0034-0.0068}^{+0.0034+0.0068}$ & $0.9690_{-0.0033-0.0065}^{+0.0034+0.0067}$ \\

$\ln(10^{10} A_\mathrm{s})$ & $3.042_{-0.014-0.027}^{+0.014+0.029}$ & $3.048_{-0.015-0.029}^{+0.014+0.031}$ & $3.050_{-0.014-0.028}^{+0.014+0.029}$ & $3.050_{-0.016-0.028}^{+0.014+0.031}$ & $3.049_{-0.015-0.027}^{+0.014+0.029}$ & $3.049_{-0.016-0.027}^{+0.014+0.030}$ \\

$\delta$ & $0.34_{-0.11}^{+0.24}<0.57$ & $0.15_{-0.12}^{+0.062}<0.31$ & $<0.055<0.12$ & $<0.073<0.15$ & $<0.044<0.097$ & $<0.079<0.16$ \\

$\Omega_\mathrm{m}$ & $0.2835_{-0.020-0.027}^{+0.013+0.032}$ & $0.2924_{-0.0054-0.012}^{+0.0067+0.011}$ & $0.3001_{-0.0042-0.0082}^{+0.0041+0.0080}$ & $0.2988_{-0.0043-0.0090}^{+0.0044+0.0083}$ & $0.3009_{-0.0039-0.0076}^{+0.0039+0.0077}$ & $0.2982_{-0.0043-0.0093}^{+0.0044+0.0086}$ \\

$\sigma_8$ & $0.840_{-0.015-0.034}^{+0.019+0.032}$ & $0.822_{-0.012-0.020}^{+0.009+0.022}$ & $0.812_{-0.007-0.013}^{+0.007+0.014}$ & $0.814_{-0.008-0.014}^{+0.006+0.016}$ & $0.811_{-0.007-0.013}^{+0.006+0.014}$ & $0.814_{-0.009-0.015}^{+0.007+0.016}$ \\

$H_0$ [km/s/Mpc] & $70.9_{-1.8-4.1}^{+2.2+3.7}$ & $69.60_{-0.88-1.4}^{+0.60+1.5}$ & $68.63_{-0.44-0.75}^{+0.34+0.80}$ & $68.78_{-0.51-0.86}^{+0.37+0.94}$ & $68.54_{-0.38-0.66}^{+0.33+0.74}$ & $68.84_{-0.52-0.88}^{+0.38+0.95}$ \\

$S_8$ & $0.816_{-0.015-0.033}^{+0.017+0.036}$ & $0.812_{-0.008-0.017}^{+0.008+0.017}$ & $0.812_{-0.009-0.017}^{+0.008+0.016}$ & $0.812_{-0.009-0.016}^{+0.008+0.017}$ & $0.812_{-0.008-0.017}^{+0.008+0.016}$ & $0.811_{-0.008-0.017}^{+0.008+0.017}$ \\

$r_{\rm{drag}}$ [Mpc] & $147.18_{-0.27-0.53}^{+0.27+0.52}$ & $147.39_{-0.21-0.43}^{+0.22+0.42}$ & $147.45_{-0.19-0.39}^{+0.20+0.38}$ & $147.45_{-0.20-0.39}^{+0.19+0.40}$ & $147.45_{-0.19-0.39}^{+0.19+0.38}$ & $147.45_{-0.20-0.40}^{+0.20+0.40}$ \\

\hline

$\Delta\chi^2_{\rm min}$ & $-0.8$ & $-0.7$ & $1.2$ & $1.4$ & $1.5$ & $1.2$  \\

$\ln \mathcal{B}_{ij}$ & $-0.4$ & $-1.4$ & $-3.3$ & $-2.7$ & $-3.5$ & $-2.7$  \\		
			
            \end{tabular}}
             \scalebox{0.85}{
			\begin{tabular}{cccccccc}
			\hline
			Parameters & CMB & CMB+DESI & CMB+DESI+PantheonPlus & CMB+DESI+Union3 & CMB+DESI+DES-Dovekie  & CMB+DESI+Union3.1 \\ \hline
					
			$\Omega_\mathrm{b} h^2$ & $0.02249_{-0.00017-0.00033}^{+0.00017+0.00034}$ & $0.02242_{-0.00013-0.00027}^{+0.00014+0.00026}$ & $0.02242_{-0.00013-0.00027}^{+0.00013+0.00025}$ & $0.02242_{-0.00013-0.00027}^{+0.00014+0.00025}$ & $0.02241_{-0.00014-0.00027}^{+0.00014+0.00027}$ & $0.02242_{-0.00013-0.00028}^{+0.00015+0.00027}$ \\

$\Omega_\mathrm{c} h^2$ & $0.1186_{-0.0015-0.0029}^{+0.0015+0.0028}$ & $0.11923_{-0.00087-0.0017}^{+0.00085+0.0017}$ & $0.11931_{-0.00087-0.0017}^{+0.00083+0.0018}$ & $0.11924_{-0.00089-0.0016}^{+0.00080+0.0018}$ & $0.11934_{-0.00091-0.0018}^{+0.00090+0.0019}$ & $0.11917_{-0.00087-0.0018}^{+0.00085+0.0017}$ \\

$100\theta_\mathrm{MC}$ & $1.04195_{-0.00030-0.00060}^{+0.00031+0.00060}$ & $1.04189_{-0.00028-0.00054}^{+0.00029+0.00053}$ & $1.04190_{-0.00026-0.00052}^{+0.00026+0.00057}$ & $1.04188_{-0.00028-0.00058}^{+0.00029+0.00055}$ & $1.04192_{-0.00028-0.00055}^{+0.00027+0.00056}$ & $1.04190_{-0.00028-0.00056}^{+0.00028+0.00054}$ \\

$\tau$ & $0.0508_{-0.0077-0.016}^{+0.0078+0.016}$ & $0.0523_{-0.0077-0.017}^{+0.0078+0.016}$ & $0.0524_{-0.0077-0.016}^{+0.0078+0.015}$ & $0.0528_{-0.0078-0.016}^{+0.0075+0.015}$ & $0.0516_{-0.0071-0.017}^{+0.0072+0.016}$ & $0.0522_{-0.0079-0.017}^{+0.0083+0.016}$ \\

$n_\mathrm{s}$ & $0.9681_{-0.0047-0.0091}^{+0.0046+0.0096}$ & $0.9666_{-0.0036-0.0074}^{+0.0037+0.0072}$ & $0.9662_{-0.0036-0.0074}^{+0.0037+0.0071}$ & $0.9665_{-0.0037-0.0072}^{+0.0035+0.0072}$ & $0.9663_{-0.0035-0.0072}^{+0.0035+0.0073}$ & $0.9666_{-0.0036-0.0069}^{+0.0036+0.0072}$ \\

$\ln(10^{10} A_\mathrm{s})$ & $3.033_{-0.016-0.033}^{+0.016+0.032}$ & $3.038_{-0.015-0.031}^{+0.015+0.032}$ & $3.039_{-0.015-0.030}^{+0.015+0.030}$ & $3.039_{-0.015-0.029}^{+0.015+0.031}$ & $3.037_{-0.014-0.032}^{+0.015+0.030}$ & $3.037_{-0.015-0.033}^{+0.017+0.031}$ \\

\rowcolor{LightSkyBlue}
$\sum m_{\nu,{\rm eff}}$ & $-0.12_{-0.15-0.34}^{+0.17+0.31}$ & $-0.053_{-0.072-0.12}^{+0.064+0.13}$ & $-0.063_{-0.064-0.11}^{+0.057+0.12}$ & $-0.060_{-0.072-0.12}^{+0.062+0.13}$ & $-0.059_{-0.069-0.12}^{+0.062+0.13}$ & $-0.060_{-0.070-0.12}^{+0.061+0.12}$ \\

$\delta$ & $0.32_{-0.13-0.31}^{+0.23+0.25}$ & $<0.11<0.24$ & $<0.034<0.083$ & $<0.044<0.11$ & $<0.032<0.073$ & $<0.046<0.11$ \\

$\Omega_\mathrm{m}$ & $0.262_{-0.025-0.052}^{+0.025+0.051}$ & $0.2911_{-0.0052-0.011}^{+0.0057+0.010}$ & $0.2960_{-0.0044-0.0087}^{+0.0043+0.0090}$ & $0.2951_{-0.0045-0.0094}^{+0.0048+0.0088}$ & $0.2967_{-0.0045-0.0089}^{+0.0044+0.0089}$ & $0.2946_{-0.0048-0.0093}^{+0.0049+0.0092}$ \\

$\sigma_8$ & $0.873_{-0.038-0.068}^{+0.032+0.078}$ & $0.838_{-0.013-0.025}^{+0.013+0.028}$ & $0.834_{-0.013-0.023}^{+0.012+0.025}$ & $0.835_{-0.014-0.028}^{+0.014+0.027}$ & $0.833_{-0.014-0.025}^{+0.013+0.026}$ & $0.834_{-0.013-0.025}^{+0.013+0.027}$ \\

$H_0$ [km/s/Mpc] & $73.0_{-3.1-5.8}^{+2.5+6.1}$ & $69.60_{-0.71-1.2}^{+0.49+1.3}$ & $69.00_{-0.39-0.78}^{+0.39+0.78}$ & $69.09_{-0.46-0.84}^{+0.39+0.89}$ & $68.93_{-0.41-0.76}^{+0.38+0.80}$ & $69.15_{-0.47-0.83}^{+0.42+0.92}$ \\

$S_8$ & $0.813_{-0.014-0.034}^{+0.017+0.034}$ & $0.826_{-0.012-0.022}^{+0.011+0.023}$ & $0.829_{-0.012-0.021}^{+0.011+0.023}$ & $0.828_{-0.012-0.023}^{+0.012+0.024}$ & $0.829_{-0.012-0.023}^{+0.012+0.023}$ & $0.827_{-0.011-0.022}^{+0.011+0.023}$ \\

$r_{\rm{drag}}$ [Mpc] & $147.37_{-0.33-0.60}^{+0.30+0.63}$ & $147.26_{-0.21-0.44}^{+0.21+0.42}$ & $147.24_{-0.22-0.41}^{+0.22+0.42}$ & $147.26_{-0.22-0.43}^{+0.22+0.42}$ & $147.24_{-0.23-0.46}^{+0.23+0.44}$ & $147.27_{-0.22-0.42}^{+0.21+0.45}$ \\

\hline

$\Delta\chi^2_{\rm min}$ & $-1.1$ & $1.8$ & $1.7$ & $2.2$ & $2.2$ & $2.2$  \\

$\ln \mathcal{B}_{ij}$ & $-0.3$ & $-2.5$ & $-3.6$ & $-3.4$ & $-4.0$ & $-3.4$  \\
				
                \hline
			                                                   
		\end{tabular}}                                                       
		\caption{ 
Summary of the observational constraints on various free and derived cosmological parameters of 
{\bf GEDE}, at 68\% and 95\% CL using several observational datasets. When $\delta$ is not well constrained, we quote its upper limits at 68\% and 95\% CL. 
The upper half of the table corresponds to the scenario with neutrino mass fixed, while the lower half of the table corresponds to the scenario including neutrino mass as a free parameter.  
}
		\label{table:gede}                   
	\end{table*}                                     
\end{center}
\endgroup

\subsection{Generalized Thawers}

Tables~\ref{table:generalized-thawing-p-1} ($p=1$),~\ref{table:generalized-thawing-p-5} ($p=5$),~\ref{table:generalized-thawing-p-15} ($p=15$) summarize the constraints on the generalized thawing DE models at 68\% and 95\% CL. The upper half of each table corresponds to the constraints for a fixed $\sum m_{\nu,{\rm eff}}$, while the lower half of the table corresponds to the free-to-vary $\sum m_{\nu,{\rm eff}}$ case.

We begin with the first generalized thawing DE model ($p=1$), with results in 
Table~\ref{table:generalized-thawing-p-1}. 
As expected~\cite{Linder:2015zxa}, the $p=1$ 
algebraic thawer gives results virtually identical to the usual (calibrated) thawer of the previous subsection, as both excellently match standard thawing scalar fields. This holds for both the fixed and varying neutrino mass cases, and for all data combinations, providing a useful cross-check of the code.

For the two scenarios with $p=5$ and $p=15$, the probability contours exhibit an approximately $2\sigma$ shift toward negative neutrino mass values.
As $p$ increases, the mean value of $w_0$ increases as well, since the steeper evolution means $w(z)$ departs from $-1$ only at lower and lower redshift, and in compensation $w_0$ rises to better fit the data. Indeed, higher $p$ improves the fit to observations (except for CMB alone). Some of the CMB+DESI+SN data combinations reach $\Delta\chi^2\lesssim-8$ and the Bayesian evidence achieves the moderate preference regime. Thus high-$p$ generalized thawers can provide good fits to survey observations, but still have the issue that $\mnu<0$ at $\approx95\%$ CL 
(and the fine tuning issues discussed in Sec.~\ref{sec-2}).

\subsection{Mirage}

Table~\ref{table:mirage} summarizes the constraints on mirage DE at 68\% and 95\% CL. Again, the upper half of the table corresponds to the constraints for a fixed $\sum m_{\nu,{\rm eff}}$, and the lower half of the table corresponds to the free-to-vary $\sum m_{\nu,{\rm eff}}$ case. Mirage DE gives results different from the other cases, as 
clearly evident from Figure~\ref{fig:2D}.

Now, roughly half of the neutrino mass posterior lies at positive values. 
Furthermore, both $\Delta \chi^2_{\rm min}$ and the Bayesian evidence analysis support this scenario over $\Lambda$CDM, at least at 
a modest level. Given that positive neutrino 
mass is consistent with the results, one could 
decide to fix $\mnu=0.06$ eV as is done in many standard analyses. In that case, the fit 
reaches $\Delta\chi^2\approx-11$ and 
$\ln B_{ij} \approx4$.

\subsection{GEDE}

In GEDE, any deviation from $\delta=0$ indicates 
a deviation from $w=-1$, with $\delta>0$ giving 
$w(z)<-1$ at all redshifts. This decreases the dark 
energy density at higher redshift for the same current density. 
In Table~\ref{table:gede} we summarize the constraints on this scenario (considering both versions: one with a fixed total neutrino mass and the other with a free-to-vary total neutrino mass). 

For none of the cases or data combinations (except CMB alone) is any significant evidence for $\delta \neq 0$ found. Hence there is 
no real distinction from $\Lambda$CDM and 
no preference (and some penalty) in model comparison from either $\Delta\chi^2$ or Bayesian evidence. This is not surprising since GEDE does not 
cross $w=-1$, and so the best the data can do 
is push it as close to $-1$ as reasonable.

However, the behavior of neutrino mass offers 
a clue. While the neutrino mass posterior still 
prefers negative values, this is now only at the 
$\approx1\sigma$ level. 
Figure~\ref{fig:2D-gede} makes this clearer. 
As $\delta$ increases, lowering the high-redshift 
DE equation of state and thus the high-redshift 
DE density, there is room for positive neutrino 
mass to contribute to the cosmic expansion at 
redshifts $z\gtrsim1$, as the data prefer. 
This is more evident in the CMB+DESI contour, 
before the SN data pull back in the direction 
of less negative $w(z)$ at lower redshift, 
forcing $\delta$ toward 0. 

So while GEDE is not a cosmological 
model preferred over $\Lambda$CDM, it does 
teach a valuable lesson regarding the physical 
characteristic allowing positive neutrino mass 
to be more prevalent in the posterior.

\section{Summary and Conclusions}
\label{sec-summary} 

Two unexpected properties of recent cosmological data are 
that they prefer the dark energy equation of state crossing 
$w=-1$ and, in certain models, they prefer the sum of neutrino 
masses to be negative, $\mnu<0$. The $w_0w_a$ cosmology does 
alleviate the negative neutrino mass issue, and one question 
is identifying which properties lead to this. To restrict the 
possibility that parameter space degeneracy is a main factor, 
we consider only single-parameter dark energy equations of 
state. Each one has a particular physical characteristic to 
help narrow down the conditions enabling positive neutrino mass. 

Generalized thawing models by construction never cross 
$w=-1$, and focus on the rapid evolution of the dark energy 
equation of state at $z\lesssim0.5$. They can closely mimic 
scalar field models ranging from the thawing class to 
PNGB or hilltop models. GEDE also never crosses $w=-1$ but 
remains phantom, and so has reduced dark energy density 
(relative to $\Lambda$CDM) at $z\gtrsim1$. The calibrated 
thawing model ($w_a=-1.58(1+w_0)$) mathematically crosses 
$w=-1$ but gives observables virtually identical to scalar 
field thawers, and hence to $p=1$ generalized thawers. Finally, 
the mirage model strongly crosses $w=-1$ and has reduced 
dark energy density at high redshift, acting very much like 
the best-fit $w_0w_a$ cosmology but with only one free parameter. 

Our results show that the key dark energy property for 
enabling positive neutrino mass is the reduction of 
dark energy density at $z\gtrsim1$. Both the mirage and 
GEDE cases exhibit this, while none of the others do. 
That the one-parameter mirage dark energy works as well as 
$w_0w_a$ indicates that opening the neutrino mass 
posterior to positive values is not due to an overly large 
parameter space. We also see that neither crossing $w=-1$ 
per se nor rapid low-redshift evolution is the cause. 
Physically, the reduced dark energy density at $z\gtrsim1$ 
due to the (actual) phantom nature there lowers the expansion 
rate $H(z\gtrsim1)$ and thus provides room for positive neutrino 
mass. A non-phantom behavior at $z\gtrsim1$ leaves no space 
for positive neutrino mass, pushing the posterior toward negative 
values despite the physical properties of neutrinos. 

That said, phantom dark energy that does not cross $w=-1$, 
such as GEDE, does not provide as good a fit to the data 
(in terms of $\Delta\chi^2$ or evidence) as dark energy that does 
cross $w=-1$ (such as mirage, or $w_0w_a$). 
Positive neutrino 
mass therefore appears to carry with it the requirement for a phantom 
crossing to fit current cosmological data as well. Note 
that all these conclusions are qualitatively unchanged 
regardless of which supernova compilation is included, or 
even when considering only DESI+CMB without supernovae.

\acknowledgments 
We thank David Rubin for providing the Union3.1 data, and Taylor Hoyt for discussions of it. 
WY has been supported by the National Natural Science Foundation of China under Grant Nos. 12547110 and 12175096. EDV is supported by a Royal Society Dorothy Hodgkin Research Fellowship.  SP acknowledges the partial support from the Department of Science and Technology (DST), Govt.~of India under the Scheme  ``Fund for Improvement of S\&T Infrastructure (FIST)'' (File No. SR/FST/MS-I/2019/41).  This article is based upon work from COST Action CA21136 Addressing observational tensions in cosmology with systematics and fundamental physics (CosmoVerse) supported by COST (European Cooperation in Science and Technology).


\bibliography{biblio}

\end{document}